\documentstyle[psfig]{mn}
\newcommand\etal{et al. }
\newcommand\refer{\par \noindent\hangindent=3pc \hangafter=1}
\newcommand\ros{{\it ROSAT}}
\newcommand\exo{{\it EXOSAT\ }}
\newcommand\ginga{{\it GINGA\ }}
\newcommand\AS{{\it ASCA\ }}
\newcommand\IRAS{{\it IRAS\ }}
\newcommand\Mrk{Markarian 766}
\newcommand\eg{e.g. }
\newcommand\etc{etc. }
\newcommand{\nH}{{N}_{\mbox{\tiny\rm H}}}
\newcommand\ie{i.e. }
\newcommand\xspec{XSPEC}
\newcommand{\half}{\mbox{H$\alpha$\ }}
\newcommand{\hbeta}{\mbox{H$\beta$\ }}

\title{Soft X--ray spectral variations of the narrow line Seyfert 1 
galaxy Markarian 766}
\author[Page \etal]{M.\,J. Page\(^{1}\), F.\,J.
Carrera\(^{1,2}\), J.P.D. Mittaz\(^{1}\), K.O. Mason\(^{1}\)\\
\(^{1}\)Mullard Space Science Laboratory, University College London,
Holmbury St Mary, Dorking, Surrey RH5 6NT, UK.\\
\(^{2}\)Instituto de F\'\i sica de Cantabria (Consejo Superior de
Investigaciones Cient\'\i ficas--Universidad de Cantabria), 39005
Santander, Spain.}

\date{}

\begin{document}
\maketitle

\begin{abstract}

The X--ray variability of the  narrow line Seyfert 1 galaxy \Mrk\ is
studied using 9 \ros\ PSPC datasets. The  
spectrum is well described by a power law combined with a
blackbody (kT \(\sim\) 70 eV) soft excess.  
Examination of flux ratio 
changes and variability amplitude in 3 X--ray bands, shows that 
the power law component varies continuously on time-scales of \(\sim 5000s\)
and is steeper when it is brighter. In contrast, variability of the
soft excess is {\it not} detected. Spectral modeling of 31 spectra from
different observations and at a range of count rates is also consistent with a
picture in which the power law is steeper when it is brighter, and in which the
soft excess component does not vary.
The power law 
variability can be explained if the power law is produced by variable 
thermal or
non-thermal Comptonization of soft photons.
This behaviour is similar to that of Galactic black hole
candidates in the low state.
The X--ray and multiwavelength properties of \Mrk\  are shown to be very
similar to those of
other
narrow line Seyfert 1s. This may mean that the rapid X--ray variability seen
in other narrow line Seyfert 1s may also not originate in their 
strong soft excess
components.

\end{abstract}

\begin{keywords}
X--rays:Seyferts - variation:spectral - variation:rapid
\end{keywords}

\section{Introduction}

Studying the  X--ray emission from active galactic nuclei allows us to probe
deep into their central regions, at scales of a few light minutes to a few 
light days from the central
engine. 
Timing and spectroscopy of the X--ray emission therefore allow us to
examine the conditions and environment in which matter accretes onto a
massive black hole.

Of particular importance is the soft X--ray band (0.1 -- 2 keV) which has
been made accessible by \ros. Two principal emission 
components are found 
in AGN spectra in this band, a power law which extends to much 
higher energies (\(>40\) keV)
and an excess at low energies 
(typically \(<0.5\) keV). This soft excess may be the
high energy tail of the big blue bump, which dominates the bolometric
luminosity of most radio quiet AGN 
(\eg Walter \& Fink 1993, Turner \& Pounds 1989). 
Understanding the
mechanism by which the soft excess is produced is therefore essential for
the construction of a realistic model of the central regions of AGN.

\subsection{Models of the X--ray Power Law and Soft Excess}

\label{sec:models}
In radio loud AGN, the mechanism responsible for the hard power law 
is probably synchrotron or synchrotron self Compton radiation
from relativistic jets (Shastri \etal 1993).
This may not be the case for the power law component seen in
the X--ray spectra of radio quiet AGN, which are in general softer than
those from  radio loud objects (Lawson \etal 1992). 
At \(>\)5 keV, radio quiet AGN show spectral features
(Fe line emission 
and a hardening of the spectrum above 10 keV, Pounds \etal 1990)
which suggest that some of the X--rays are reflected from
material near the black hole; 
in MCG-6-30-15 and other Seyfert 1s for which high signal to noise 
\AS data are available, 
the Fe line is relativistically broadened (Nandra \etal 1997). 
Therefore, the X--ray power law component 
may originate in a hot corona above a relativistic accretion disk so that
\(\sim
50\%\) of the X--rays generated in the corona shine down on the
accretion disk, and are reflected from its surface. In this model the X--ray
power law may be due to photons which are emitted originally from
the accretion disk itself,  and then
Compton upscattered either by electrons in a hot
plasma (Walter \& Courvoisier 1992), or by a non-thermal distribution of
relativistic electrons injected into a compact region, in which the
radiation density may be so high that photon - photon collisions result in
production of electron - positron pairs (Zdziarsky \etal 1990).

A number of models have been put forward to explain the soft excess emission
of AGN. One obvious model is that the soft excess (and the whole of the big
blue bump) is thermal emission from an accretion disk surrounding a
supermassive black hole (\eg Ross, Fabian and Mineshige 1992). 
Alternatively, the soft excess may be
due to reprocessing of the power law component by 
optically thin clouds that surround the central
regions (\eg Guilbert \& Rees 1988), 
or the surface of 
an accretion disk (\eg Ross \& Fabian 1993). Reprocessing of the radiation
by electron - positron pairs can also produce a soft excess (Zdziarski \etal
1990).
In the reprocessing models, the soft excess responds to changes
in the power law flux, with a time delay
related to the distance between 
the power law emitter and the reprocessing material. 
If the power law component is caused by Compton upscattering of soft X--ray and
UV accretion disk
photons, and accretion disk instabilities drive the variability,
then the delay between the power law and soft
excess components is expected to be in the opposite sense, i.e. the soft
excess changes before the power law component.
If the soft excess is due to reprocessing of X--rays by electron -
positron pairs, time delays between the power law component and soft
excess depend on the compactness of the emission region (Done \& Fabian 1989).

\subsection{The Narrow Line Seyfert 1 Galaxy \Mrk}

Narrow line Seyfert 1 galaxies (NLS1s hereafter) 
are known to have strong soft excesses, 
(Puchnarewicz \etal 1992, Boller, Brandt \& Fink 1996)
and are therefore  obvious candidates for investigating the interrelation
between the power law and soft excess components.
However, detailed studies of individual objects have tended to concentrate
on those with extreme properties, of either abnormally steep soft X--ray
spectra and/or unusually large variability, \eg RE J1034+393 
(Puchnarewicz \etal 1995), IRAS 13224-3809
(Otani \etal 1996), RE J1237+264 (Brandt, Pounds \& Fink 1995).

In this paper we present a study of the X--ray emission of the nearby
(z=0.013), radio quiet, NLS1 galaxy \Mrk. \Mrk\  fell within the field of view
of the \ros\ PSPC during 9 different observations, was
observed at least once in each of the years 1991 to 1994 inclusive, and is
a bright PSPC source (1 -- 6 counts s\(^{-1}\)). The long sampling
time scale
provides a rare opportunity 
to study {\it both} the long and short term behaviour of the
source.

\Mrk\  has a significant but not {\it extreme} soft excess (Boller, Brandt
\& Fink 1996), and is known to vary in a few hours 
(Molendi, Maccacaro \& Schaeidt 1993); these are essential properties if
the relations between the soft excess and power law components are to be
investigated within the time scales of individual \ros\ observations. 
It is a strong Fe II emitter (Gonz\'alez-Delgado \& P\'erez 1996)
and has significant optical polarization of \(>2\%\) perpendicular to its
radio axis (Goodrich 1989). 

Results from some \ros\ observations of \Mrk\  have already been
published. Molendi, Maccacaro \& Schaeidt (1993) showed that \Mrk\ 
varied by a factor of \(\sim 3\) in the two days of observation during the
\ros\ all sky survey. 
Molendi \& Maccacaro (1994) and Netzer, Turner \& George (1994) used pointed
\ros\ observations and 
concluded that the variable X--ray spectrum is not the result of a
variable warm absorber. 
\Mrk\   has recently been observed with \AS (Leighly \etal 1996), and
was shown to have a warm absorber
(evidenced by absorption edges at \(\sim 0.7 - 0.8\) keV) and a 
variable power law component. Although a soft
excess was detected, it is  poorly constrained because \AS is not sensitive
to photons with energies \(<0.4\) keV.

\subsection{The Structure of this Paper}

The 9 \ros\ observations and details of the data reduction are presented in
Section \ref{sec:observations}. 
The \ros\ data are examined in two ways: 
using data from three broad X--ray bands 
and spectral fitting at full resolution.
Together they provide the maximum information about the spectral 
variability of \Mrk\ via the following procedure:

First, X--ray lightcurves are constructed in three energy bands (Section
\ref{sec:lightcurves}). 
Using the lightcurves as a guide,
X--ray spectra are then constructed at six different flux levels (Section
\ref{sec:spectral}), allowing us to 
examine the spectral shape of \Mrk\  as it changes in brightness, and 
determine the contributions of the different emission components to 
the three X--ray bands.
Variations in the three colour lightcurves and hardness ratios are then 
used to study variability of the emission components (Section
\ref{sec:colvar}) in a relatively model independent fashion. 
Finally, we cross correlate the flux in the three different 
bands to look for temporal relations between the emission at different energies
(Section \ref{sec:crosscorrelation}).

Our results are discussed in Section \ref{sec:discussion}, and the X--ray
and multiwavelength properties of \Mrk\ are compared to those of other NLS1s
in Section \ref{sec:others}. Our conclusions are presented in Section
\ref{sec:conclusions}. We discuss the effect of the 
PSPC gain drift on the \Mrk\ data in Appendix A.  

\section{Observations}
\label{sec:observations}

\begin{table*}
\caption{The 9 \ros\ observation datasets used in this analysis}
\label{tab:observations}
\begin{tabular}{l@{\hspace{2mm}}lcccccc}

 Name & ROR & start & start &observation& exposure  
 & off-axis & count rate  \\
&number&date&MJD&length&time&angle&\\
&&& & (seconds) & (seconds) & (arcmin) & counts/s \\ 
\hline
  & & & & & & & \\
P1 & RP700221n00 & 15 June 1991 & 48422.0 & 86246  & 8313  & 19.4 *& 1.1 \\
P2 & RP701203n00 & 16 June 1992 & 48789.2 & 426470 & 6563  & 0.0   & 3.8 \\
P3 & RP701056n00 & 18 June 1992 & 48791.1 & 419862 & 7178  & 43.9 *& 2.2 \\
P4 & RP701091n00 & 8 Dec. 1992  & 48964.6 & 138240 & 5871  & 0.0   & 3.5 \\
P5 & RP701203m01 & 9 Dec. 1992  & 48965.1 & 8640   & 2597  & 39.6  & 3.8 \\
P6 & RP700970n00 & 21 Dec. 1992 & 48977.7 & 149472 & 14613 & 13.5  & 3.3 \\
P7 & RP701413n00 & 16 Dec. 1993 & 49337.5 & 25920  & 3093  & 39.0  & 3.1 \\
P8 & RP701353n00 & 17 Dec. 1993 & 49338.9 & 17280  & 3044  & 0.0   & 2.3 \\
P9 & RP701091a01 & 5 June 1994  & 49508.8 & 7600   & 2703  & 0.0   & 6.3 \\
&&&&&&&\\
\multicolumn{8}{l}{* In these observations \Mrk\  is close to the PSPC ribs}
\end{tabular}
\end{table*}

Details of the 9 \ros\ datasets used in this analysis are given in Table
\ref{tab:observations}.
We will refer to them as P1, P2, \etc
All data were are REV2 data obtained using PSPC-B. All data were
processed with the FTOOLS PCPICOR calibration task; all subsequent 
data reduction has been carried
out using the Starlink ASTERIX package.
The observations
have been filtered using the same event rate and master-veto rate criteria,
and all times with poor aspect solution have been excluded.

\Mrk\  was at the centre of the PSPC field of view during 4 of the
observations. In the remaining 5 it was at 
different off-axis angles, ranging between 13.5 and 43.9 arc minutes 
(see Table \ref{tab:observations}). 

\subsection{Comparing data from different observations}
\label{sec:comparing}

Comparison of data from the different \ros\ observations should only be made 
with considerable caution for two reasons: 

\vspace{1mm}
\noindent
1) The point spread
function (PSF) changes with different off-axis angles.

\noindent
2) The PSPC gain varies with off-axis angle and with time.  

\vspace{1mm}
\noindent
The PSF dependence on off-axis angle 
is reasonably well understood and is dealt with in Section
\ref{sec:psf}.

The PSPC gain variation with time and off-axis angle, is well
documented (Prieto, Hasinger \& Snowden 1994, Snowden \etal
1995, Turner 1993) and its main effect  
is that on-axis spectra taken in the later stages of the PSPC-B
lifetime are systematically different to spectra taken off-axis and/or
earlier in the mission.
This gain variation has some impact on this work, because it means that there
is a significant difference in 
PSPC response between 
observations (P4, P8, P9) and the other observations.
 The FTOOLS task PCPICOR
uses Al K\(\alpha\) detector gain maps, interpolated in time, to correct for
the PSPC gain variations, and this has been applied to all the \ros\ data used
here. This makes a significant improvement to the data but the authors consider
that there are still some differences in response between the different observations;
this is discussed in detail in Appendix A.
However, the actual magnitude of the gain drift (\(\sim 3\%\) in the energy
scale, before correction by PCPICOR, Snowden \etal 1995)
means that it should have a very small effect on the
colour based 
hardness ratios used in this work.
Note that as well as the drift of the PSPC gain there was an  
intentional  change of PSPC gain made in October 1991.

\subsection{Use of Off-Axis Observations}
\label{offaxis}

The point
spread function (PSF) of the \ros\ PSPC changes dramatically with off-axis
angle, hence 
for the six different detector locations, different
regions have been used for collecting source and background counts. 
Source and background regions were chosen to be at similar off-axis
angles; they do not
overlap with the region obscured by the PSPC rib support structure, and 
contain no
contaminating bright sources. The use of observations with different off-axis
angles is justified because the dependence of the \ros\ PSF with off-axis
angle is well known  except for the very low energy response 
(Hasinger \etal 1994a). In observations in which \Mrk\  is far off-axis (P3, P5
and P7), large circles (6 arc min radius) 
have been used as source and background regions to ensure that the majority
of the source counts are included.

In two observations (P1 and P3), 
\Mrk\  was occulted by the ribs of the PSPC. To avoid any systematic errors
in the X--ray flux as a function of time, data from all times when the
source extraction region entered the ribs have been excluded;
in general this means that only
a portion of the spacecraft wobble period has been included (\ros\ wobbles
once every 400 seconds to prevent the detector window wire grid from
systematically occulting sources). Note that this procedure is not expected
to introduce significant spurious variability
because we used the {\it same} part of each spacecraft wobble.
The effectiveness of
this technique was verified using time filtered images and lightcurves
folded on the 400 second wobble period of \ros.

Because of the proximity of \Mrk\  to the ribs in P1 it was necessary to
use a small circle of only 0.9 arc min radius to collect source counts, and
hence the correction for counts falling outside this circle is much
larger for P1 than that for the other observations 
(see Section \ref{sec:psf}). 
P1 also differs
from the other observations in that it took place
before the October 1991 change of the gain (and hence spectral
response) of the \ros\ PSPC-B detector. 
P1 contains the most
time intensive monitoring of \Mrk, and during P1 \Mrk\  was particularly faint;
for these reasons it is a particularly interesting observation and is retained 
despite the technical difficulties.

\section{The Three X--ray Bands}
\label{sec:lightcurves}

The primary purpose for constructing a light curve in three different X--ray
bands is to examine how the overall spectrum changes with time. 
The three bands are
therefore required to have energy responses which are as independent as
possible to prevent the smearing out of spectral changes when photons of the
same energy contribute to the flux in more than one band. With a
proportional counter such as the \ros\ PSPC, it is impossible to find three
energy bands which are {\it completely} independent.
We have chosen the bands R1L, R4, and R7 (channels 11-19, 52-69 and
132-201 respectively) from Snowden \etal (1994). 
They are centred at approximately 0.2, 0.7 and 1.7 keV respectively, and their
energy responses overlap with effective areas
\(<10\%\) of their peak. Note that with bands this narrow, a
large number of PSPC channels (and hence a considerable number of source
counts) are excluded; this is acceptable because \Mrk\ is a bright source.

\subsection{Correction for the PSPC PSF}
\label{sec:psf}

It was noted in Section \ref{offaxis} that the \ros\ PSPC 
PSF changes with off-axis angle; 
some fraction of the PSF 
lies outside the circular region used to collect source counts. 
It is therefore necessary to renormalize the
count rates to the full PSF 
before data from observations
with different off-axis angles can be compared.
The PSF of the \ros\ PSPC depends on energy as well as position.
At the energies corresponding to the 
R4 and R7 bands, the dependence of the PSF with off-axis angle is well
understood, and is described by the analytic expressions given in Hasinger
\etal (1994a); 
R4 and R7 count rates have therefore been renormalized using these
expressions.

Hasinger \etal (1994a) do
not provide a good description of the PSF at low energies (\ie the R1L band)
because of a problem known as ghost imaging, in which the positions of low
pulse height events can be
incorrectly determined by the PSPC (see Snowden \etal 1994).
The counts in R1L have
been corrected using a Gaussian fit to the observed R1L 
PSF in each observation. The 90\% statistical error on this Gaussian fit
translates to a 5\% uncertainty in the corrected  R1L flux of
observation P1, which has the largest uncertainty because only a small
circle is used to collect source counts (see Section  \ref{offaxis}). In
the other observations the renormalization of R1L amounts to a change of 
less than 10\%,
 and hence the uncertainty on the R1L flux from the PSF correction 
 is not likely to be
more than a few percent. 

\subsection{Hardness Ratio Definitions}

Two hardness ratios have been used to examine spectral
changes:
\[HR_{\rm soft}=\frac{R4-R1L}{R4+R1L}\]
\[HR_{\rm hard}=\frac{R7-R4}{R7+R4}\]
Both \(HR_{\rm soft}\) and \(HR_{\rm hard}\) 
increase as the spectrum becomes harder.

\subsection{The X--ray Lightcurves}

%used /disk/xray/mjp/mkn/keep/mkn.pro, coltime.qdp
\begin{figure*}
\begin{center}
\leavevmode
\psfig{file=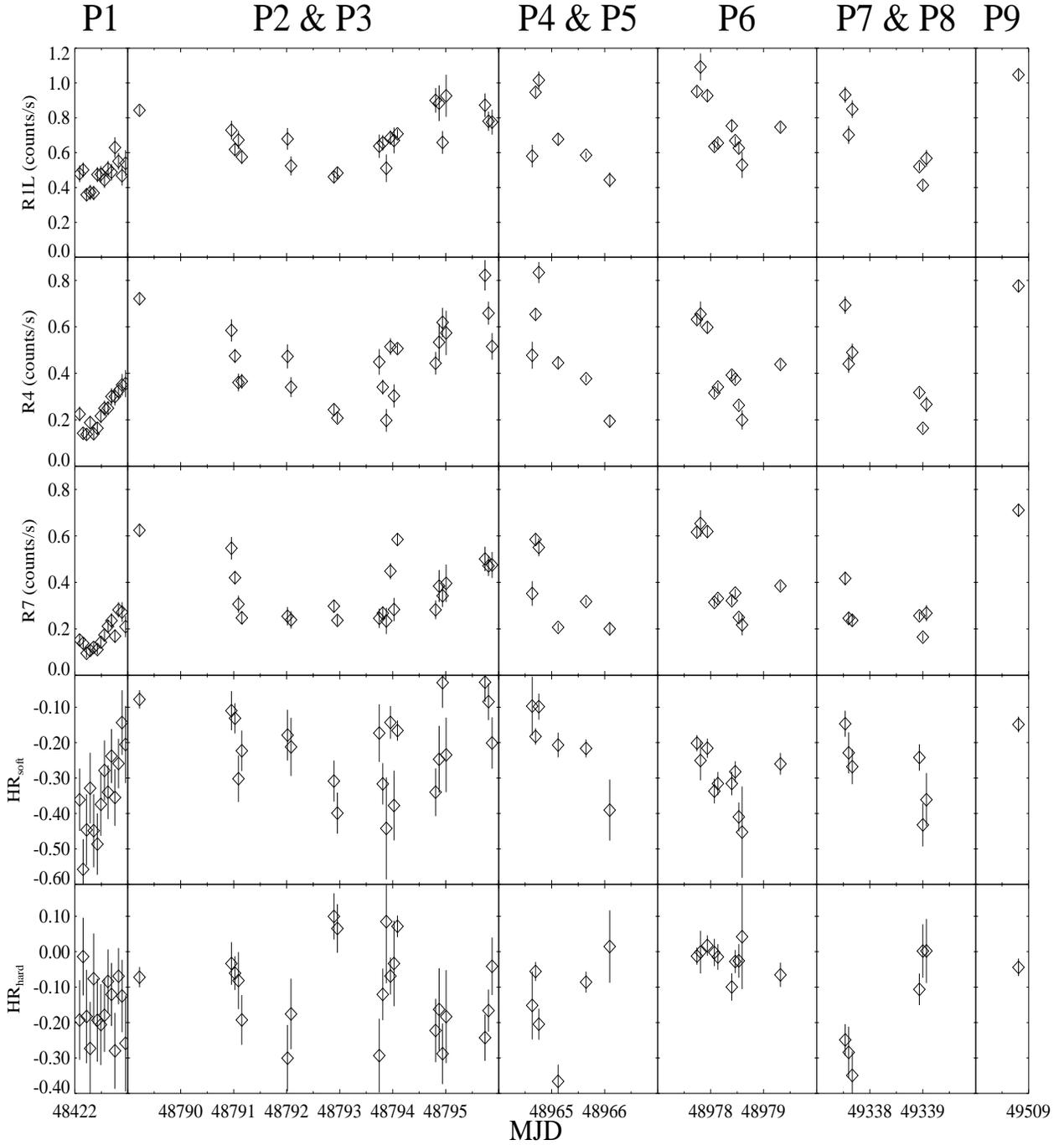,height=180mm,width=165mm}
\caption
{Variations in the 3 colour count rates and hardness
ratios of \Mrk\  binned in \ros\ orbits (5760s). 
The vertical divisions
indicate breaks
between observations; the X--axis has the same scale for all observations.}
\label{fig:lightcurves}
\end{center}
\end{figure*}

To obtain good signal to noise (typically \(>\)30), counts in each
band have been binned to 5760 seconds (i.e. 1 \ros\ orbit).

The X--ray lightcurves in the three X--ray bands for the four years of
observation are shown in the top three panels of Fig. 
\ref{fig:lightcurves}; 
note that the x-axis is not continuous. 
\Mrk\  exhibits substantial variability (\(>\) 
factor 2 in the R4 and R7 bands) 
during every observation,
excepting observations P5 and P9, which lasted only a single
\ros\ orbit each. 
Hardness ratios, computed from the three light curves are
shown in the bottom two panels of Fig. \ref{fig:lightcurves}. 
Hardness ratio changes, and
therefore changes in the spectral shape, are seen during all 
observations (again except P5 and P9 which are too short for the variability
to be addressed). The variability will be quantified later 
in Section \ref{sec:variamp} when the 3 colour and hardness ratio variability
is looked at in more detail.

\section{Spectral Modeling}
\label{sec:spectral}
\subsection{Splitting the observations}

\label{sec:splitting}

Before relating the lightcurves and hardness ratio changes to changes in
specific components of the X--ray
spectrum of \Mrk, it is necessary to examine the spectrum itself, and
obtain a good spectral model. 

The variable hardness ratios shown in Fig. \ref{fig:lightcurves}
show that the spectrum is changing within the observations. 
We have extracted several spectra from each observation, that correspond to
different levels of brightness of \Mrk, to explore the full dynamic range of
spectral shapes. 

Traditionally, the overall \ros\ count rate would be used to define the
intensity of the source (\eg Molendi and Maccacaro 1994). This is not
practical for the current dataset which includes observations at different
off-axis angles because 
the broadband \ros\ count rate is sensitive to the position of
the source on the PSPC. It  
must be convolved with a model spectrum to
correct for the proportion of the PSF outside the source extraction 
circle, because the PSF is energy dependent.
To extract spectra as a function of overall \ros\ count rate, consistently
for each observation, would therefore require a prior knowledge of the
spectra to be extracted.

This problem has been avoided by using the R4 count rate to define the
intensity of \Mrk, instead of the broadband count rate.
R4 covers a narrow enough energy range that it is
easy to correct for the PSF in a consistent way for all observations without
a detailed prior knowledge of the spectrum. R7 could equally have been used;
R4 has been preferred
because it is more central in the \ros\ bandpass.

Six different levels of intensity have been defined according to the R4
count rate, and observations have been split up by their R4
count rates averaged in 5760 second bins, as shown in Table
\ref{tab:exposures}. 
For each observation, all counts collected
during a given range of R4 count rate were combined to make a single spectrum. 
Up
to six spectra were therefore constructed from each observation; as seen in
Table \ref{tab:exposures} a total of 31 spectra were constructed.

\begin{table*}
\caption{Exposure times of spectra taken for different R4 count rates and
observations.}
\label{tab:exposures}
\scriptsize
\begin{tabular}{lcccccc}
& & & & & & \\
Data & time & time & time & time & time & time \\
& \(R4<0.2\) & \(0.2<R4<0.3\) & \(0.3<R4<0.4\) & \(0.4<R4<0.5\) &
\(0.5<R4<0.6\) & \(0.6<R4\) \\
& (seconds) & (seconds) & (seconds) & (seconds) & (seconds) & (seconds) \\
\hline
& & & & & & \\
P1 & 4132 & 2803 & 1378 & -- & -- & -- \\
P2 & -- & 1303 & 543 & 2856 & 575 & 1286 \\
P3 & -- & 216 & 2466 & 2874 & 571 & 1051 \\
P4 & 870 & -- & 343 & 189 & -- & 4469 \\
P5 & -- & -- & -- & 2597 & -- & -- \\
P6 & 165 & 1315 & 6683 & 2585 & 1502 & 2363 \\
P7 & -- & -- & 1233 & 896 & -- & 964 \\
P8 & 1186 & 1047 & 811 & -- & -- & -- \\
P9 & -- & -- & -- & -- & -- & 2703 \\
& & & & & & \\
\end{tabular}
\end{table*}

We did not combine spectra from different observations 
because of their different off-axis angles and hence effective areas, and
because of the possible response differences between the observations caused by
the PSPC temporal/spatial gain drift. 
The effective area (as a function of energy) 
for each spectrum has been normalized to the source count collecting region 
using the analytical PSF expressions of
Hasinger \etal (1994a). 

\begin{figure*}
\begin{center}
\leavevmode
\psfig{file=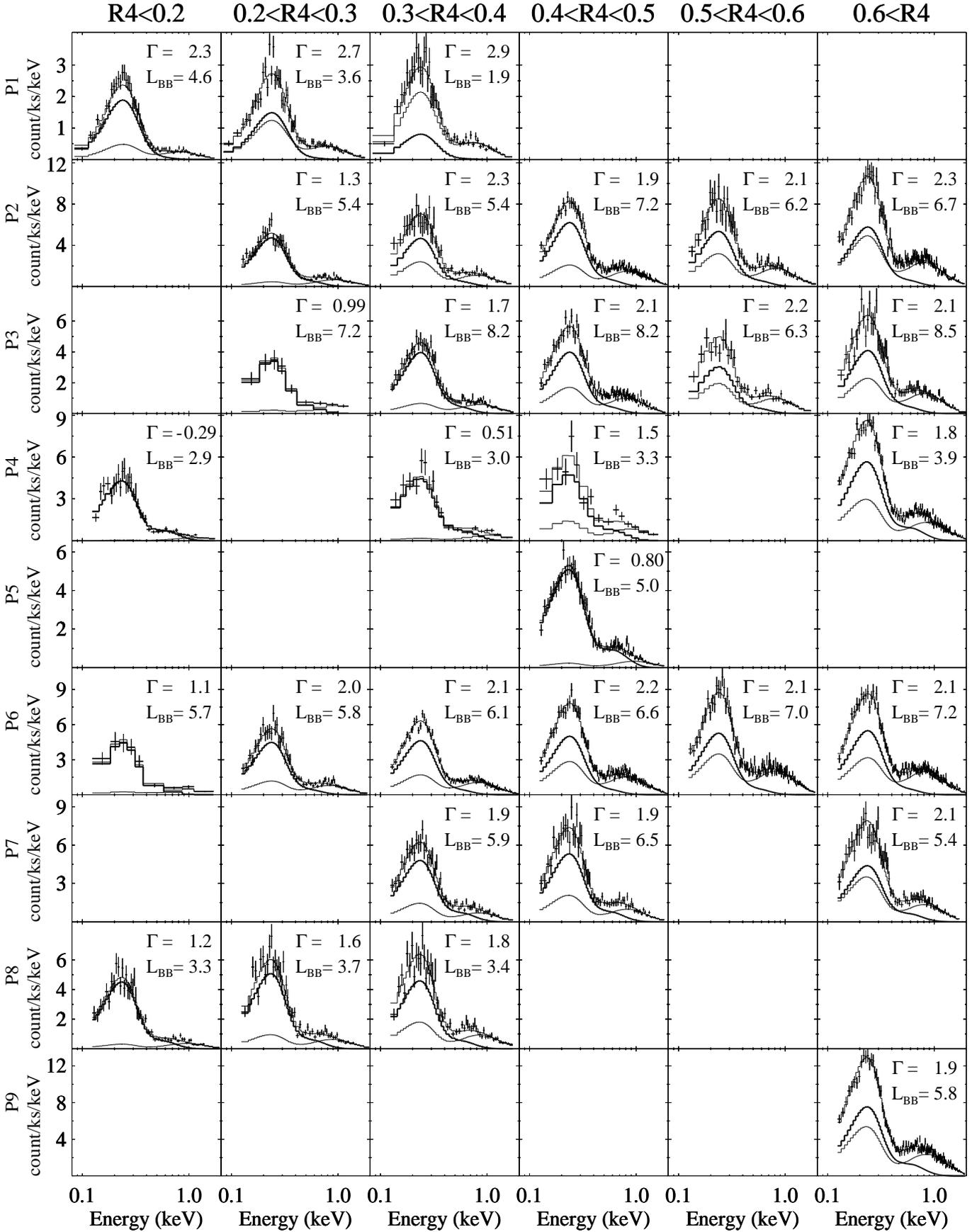,height=230mm,width=180mm}
% used to use /disk/xray/mjp/mkn/newspec/rospecomplot.pro
\caption{The 31 binned PSPC spectra (data points) with the best fitting power
law + black body soft excess models 
as stepped lines: the power law and soft excess components are shown separately
with the soft excess in bold,
as well as the combined model. 
$\Gamma$ is the photon index
of the power law and L\(_{BB}\) is the normalization
of the black body soft excess in units of 
${\rm 10^{37} erg s^{-1}}/D^{2}$ where $D$ is distance to
source in kpc. The remaining model parameters are given in row B of 
Table \ref{tab:fitting}.}
\label{fig:allspec}
\end{center}
\end{figure*}

Spectra were binned in energy with a minimum of 30 counts per bin,
and analyzed using
the spectral fitting program \xspec;
bad channels (1-7 and \(>200\) for P1, 1-11 and \(>200\) for the other
observations, as defined in Snowden \etal 1994) were ignored and an additional
systematic error of 2\% assumed for each channel to reflect the uncertainty
in the PSPC energy response.
As stated above, observation P1 was made before the PSPC gain reduction in
October 1991 and hence 
the spectra from this observation have been fitted using 
the high gain response matrix; the other observations have been fitted using
the low gain response matrix.
The 31 binned PSPC spectra are shown in Fig. \ref{fig:allspec}.

\begin{figure*}
\begin{center}
\leavevmode
\psfig{file=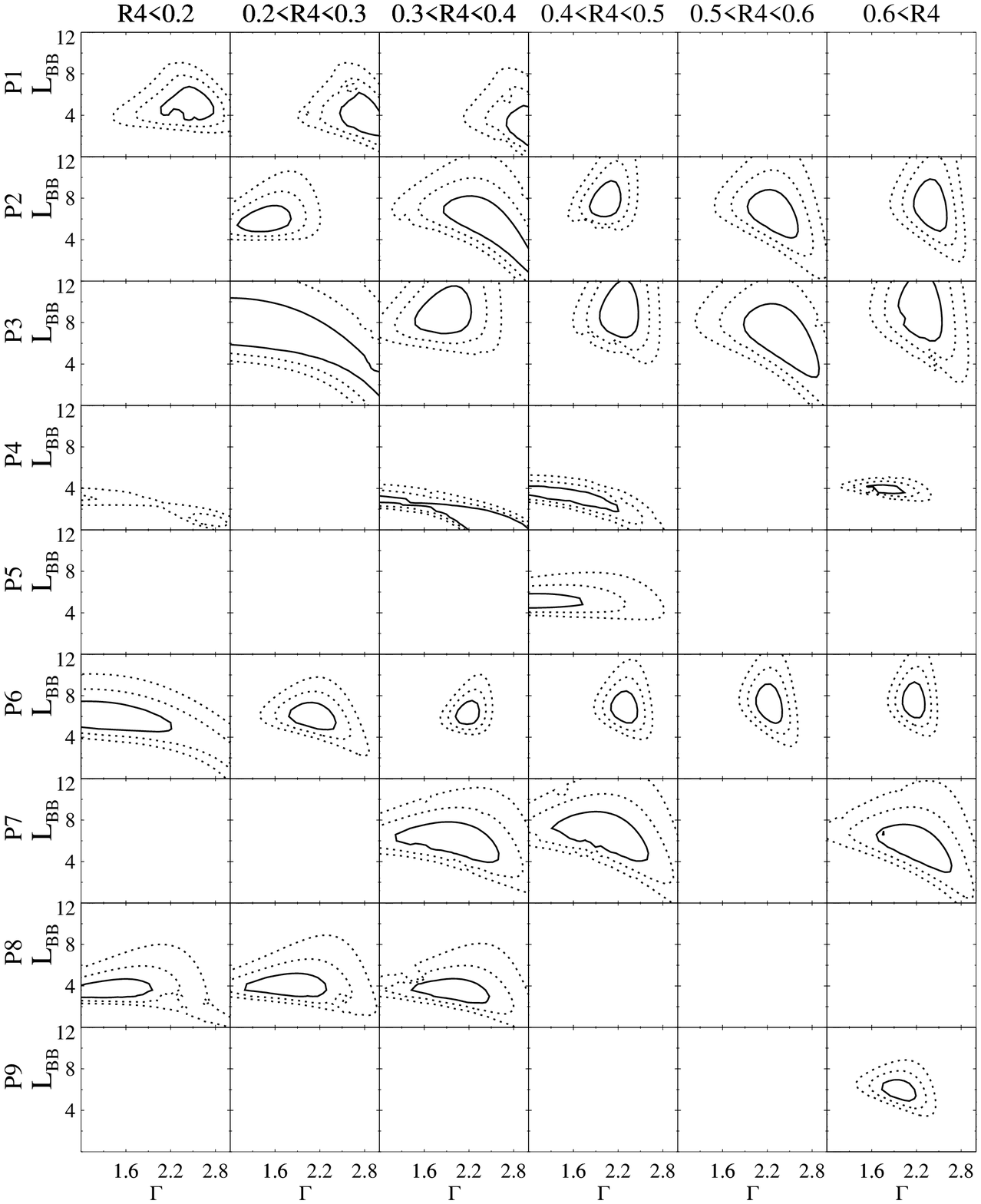,height=230mm,width=180mm}
% used to use /disk/xray/mjp/mkn/newspec/rospecont.pro
\caption{Confidence contours for the power law photon index $\Gamma$ and 
black body
soft excess normalization $L_{BB}$ (in units of 
${\rm 10^{37} erg s^{-1}}/D^{2}$  where $D$ is distance to
source in kpc). The model is the same as the one used for
Fig. \ref{fig:allspec} and 
the remaining model parameters are given in row B of 
Table \ref{tab:fitting}.}
\label{fig:allcont}
\end{center}
\end{figure*}

\subsection{Model Fitting}
\begin{figure}
\begin{center}
\leavevmode
\psfig{file=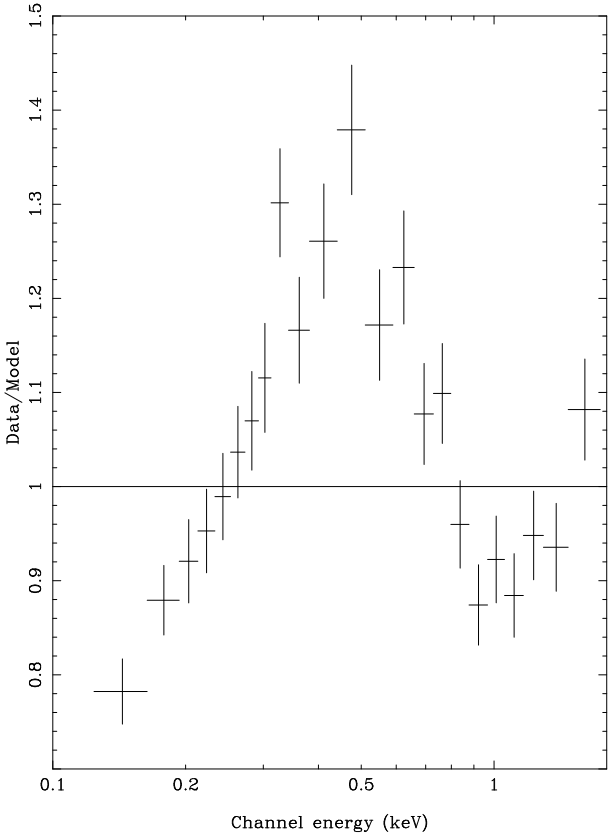,height=110mm,width=80mm}
\caption
{Ratio of the 0.4$<$R4$<$0.5 P2 spectrum to 
a power law model (best fit $\Gamma=2.2$)
with fixed Galactic $\nH$. Note the residuals at the very lowest
energies which are indicative of intrinsic neutral absorption and the bump
from 0.3 to 0.7 keV which suggests a soft excess. For clarity 
the residuals 
are shown with
larger energy bins than those used in the spectral fitting.}
\label{fig:tmprati}
\end{center}
\end{figure}

As a starting point for spectral modeling, we used a simple 
power law model with neutral absorption fixed at the Galactic
column of \(1.76\times 10^{20}\ {\rm cm^{-2}}\) 
(Stark \etal 1992). The power law
slope and normalization are fit individually for all 31 spectra. The
fitted photon indices \(\Gamma\) range between 2.1 and 2.7
but the overall fit
is very poor (\(\chi^{2}/\nu=1.9\) for 2261 degrees of freedom), 
and the residuals show evidence for
both a soft excess and additional neutral absorption (see Fig.
\ref{fig:tmprati}). 

Allowing the absorbing column to vary in each spectral fit results in a
significant improvement in the fit (\(\chi^{2}/\nu=1.3\) for 2313 degrees of
freedom) and results in a mean column which is larger than the Galactic value. 
The redshift of \Mrk\  is low enough (\(z=0.013\)),
that using a redshifted absorber to represent the additional column instead
of \(z=0\) column makes no appreciable
difference to the goodness of fit or the fitted column; this was verified 
using \xspec.

Addition of a
blackbody component
to represent the soft excess also results in a
significantly better, although still poor, 
fit (\(\chi^{2}/\nu=1.2\) for 2199 degrees of freedom) with fixed \(\nH\). 
Adding a blackbody
component {\it and} allowing the column to vary between spectra 
results in a good fit, with
\(\chi^{2}/\nu=1.0\) for 2168 degrees of freedom.

A warm (partially ionized) absorber can often 
mimic the effect of a soft excess in \ros\ spectra.
Its principal signature in the \ros\ band is the presence of 
absorption edges at 0.74  keV 
and 0.87 keV (rest energies) from O VII and O VIII respectively. These edges
can not be resolved from each other with the \ros\ PSPC, 
and can be approximated as a single
edge. It is known that \Mrk\  has a warm absorber from \AS data. 
Approximating the warm absorber as a single edge in the \AS data
resulted in an observed edge energy of 0.75 keV (Leighly \etal 1996), 
and to keep the number of
fitted parameters to a minimum we have used an edge with a fixed observed
energy of 0.75 keV to represent a warm absorber.
Substituting the blackbody soft excess with an edge at 0.75 keV, 
again leaving the neutral
column free, results in \(\chi^{2}/\nu=1.1\) for 2199 degrees of freedom.
This is better than the fit without the edge, but is much poorer than the fit
with a soft excess (the edge model can be statistically rejected with 95\%
confidence).
Including both an edge and a blackbody soft excess results in only a very
slightly better \(\chi^{2}/\nu\) than using just the blackbody soft excess; an
F-test shows that the 
improvement in the fit
is not significant, hence we do not include an edge in our spectral models 
for the remainder of this paper.

A good fit to the \ros\ spectrum of \Mrk\  therefore requires three components:
a power law, a soft excess and some intrinsic neutral absorption. 

\subsection{Absorption}

\begin{table*}
\caption
{Fitting of spectra grouped by observation. For each model,
$\chi^{2}/\nu $ is given by observation and in total. In cases where spectra
have been fitted using only a single value of one or more parameter per
observation, the parameters are listed in parentheses beneath the model
and values are given for each observation.
Errors are 68\% for one interesting parameter.}
\scriptsize
\begin{tabular}{@{\hspace{0mm}}l@{\hspace{2mm}}c@{\hspace{4mm}}c@{\hspace{4mm}}c@{\hspace{4mm}}c@{\hspace{2mm}}c@{\hspace{1mm}}c@{\hspace{2mm}}c@{\hspace{2mm}}c@{\hspace{1mm}}c@{\hspace{2mm}}c@{\hspace{1mm}}c@{\hspace{2mm}}}
&&&&&&&&&&\\
&&&&&&&&&&\\
&{\bf
Model}&P1&P2&P3&P4&P5&P6&P7&P8&P9&Total\\
&&&&&&&&&&\\
\hline
&{\bf Data} &175&417&305&225&84&612&194&158&153&2323\\
&{\bf points}&&&&&&&&&&\\
\hline
&&&&&&&&&&&\\
A&{\bf
BB}&166/162&353/396&298/284&253/208&86/79&596/587&162/181&187/145&134/147&2235/2190\\
&(\(\nH\))&2.7\(\pm\)0.4&3.2\(\pm\)0.3&3.6\(\pm\)0.3&1.9\(\pm\)0.2&2.1\(\pm\)0.4&3.5\(\pm\)0.3&2.6\(\pm\)0.4&2.3\(\pm\)0.3&2.5\(\pm\)0.3&\\
&&&&&&&&&&&\\
B&{\bf
BB}&171/164&374/400&308/288&285/211&86/79&609/592&164/183&198/147&134/148&2329/2212\\
&(\(T\))&54\(\pm\)5&68\(\pm\)4&69\(\pm\)5&91\(\pm\)5&96\(\pm\)6&65\(\pm\)5&82\(\pm\)7&77\(\pm\)5&88\(\pm\)5&\\
&(\(\nH\))&3.0\(\pm\)0.4&3.7\(\pm\)0.3&3.7\(\pm\)0.4&2.1\(\pm\)0.3&2.1\(\pm\)0.4&3.7\(\pm\)0.3&2.8\(\pm\)0.3&2.7\(\pm\)0.4&2.5\(\pm\)0.3&\\
&&&&&&&&&&&\\
C&{\bf
BR}&186/162&360/396&307/284&252/208&79/79&589/587&170/181&181/145&139/148&2254/2190\\
&(\(\nH\))&3.1\(\pm\)0.2&3.7\(\pm\)0.4&3.9\(\pm\)0.4&2.9\(\pm\)0.1&3.1\(\pm\)0.2&4.0\(\pm\)0.2&3.2\(\pm\)0.3&3.0\(\pm\)0.1&3.0\(\pm\)0.2&\\
&&&&&&&&&&&\\
D&{\bf
2BB}&163/162&350/396&297/284&276/208&93/79&586/587&164/181&188/145&138/148&2255/2190\\
&(\(\nH\))&3.1\(\pm\)0.2&3.7\(\pm\)0.4&3.9\(\pm\)0.4&2.9\(\pm\)0.1&3.1\(\pm\)0.2&4.0\(\pm\)0.2&3.2\(\pm\)0.3&3.0\(\pm\)0.1&3.0\(\pm\)0.2&\\
&&&&&&&&&&& \\
\hline
&&&&&&&&&&&\\
\multicolumn{12}{l}{{\bf BB} model spectrum has a power law, a blackbody
soft excess and absorption by neutral material}\\
\multicolumn{12}{l}{{\bf 2BB} model spectrum has a power law, two black
bodies at temperatures of 50 eV and 80 eV}\\ 
\multicolumn{12}{l}{to represent the
soft excess, and absorption by neutral material}\\
\multicolumn{12}{l}{{\bf BR} model spectrum has a power law, a bremsstrahlung 
soft excess and absorption by neutral material}\\
\multicolumn{12}{l}{$T$ is the
blackbody temperature in eV}\\
\multicolumn{12}{l}{$L_{BB}$ is the blackbody normalization in
units of
${\rm 10^{37} erg s^{-1}}/D^{2}$ where $D$ is distance to
source in kpc}\\
\multicolumn{12}{l}{$\nH$ is the neutral absorbing column
in units of $10^{20} {\rm cm^{-2}}$}\\
\label{tab:fitting}
\end{tabular}
\end{table*}
\label{sec:absorption}

In the previous section 
the neutral absorbing column was allowed to vary freely between
spectra. The short term stability of this fitted column was tested by 
fitting the 31 spectra with only a single value of
\(\nH\) for each observation (\ie 9 values of \(\nH\)), 
once using a blackbody, once using two black
bodies with fixed temperatures of 50 eV and 80 eV (see Section
\ref{sec:softshape}), and once using a
bremsstrahlung to represent the soft excess.
\(\chi^{2}\) and fitted values of \(\nH\) 
are shown in Table \ref{tab:fitting}; in each case a small
reduction in \(\chi^{2}/\nu\) is found, 
compared to that when the column is fitted
individually for each spectrum. The F-test shows that for any of these soft
excess shapes, allowing \(\nH\) to vary within each observation does not
lead to a significantly better fit than using a single value of \(\nH\) for
each observation, and when the column is allowed a different value for each
spectrum within an observation, consistent values are found. 
This shows that there is no detectable change 
in the fitted neutral absorbing
column within any of our observations, regardless of which model is used 
for the soft excess. The different fitted column values from the
different observations could be caused by the temporal/spatial gain
variation of the PSPC; see appendix A for some discussion of this.

\subsection{The Shape of the Soft Excess}

\label{sec:softshape}
Several models were tried for the soft excess component, and the lowest
\(\chi^{2}\) was found for a single blackbody 
(model A in Table \ref{tab:fitting}).
The soft excess was also parameterized as a bremsstrahlung and as two black bodies (with energies fixed at 50
and 80 eV), both of which have broader shapes than a single black body. 
The values of \(\chi^{2}\) are slightly poorer for these models (C and D in
Table \ref{tab:fitting}) but they are still acceptable fits to the data, with
\(\chi^{2}/\nu=1.0\). 

\subsection{Variability of the model components}

\label{sec:modelvar}

\begin{figure}
\begin{center}
\leavevmode
\psfig{file=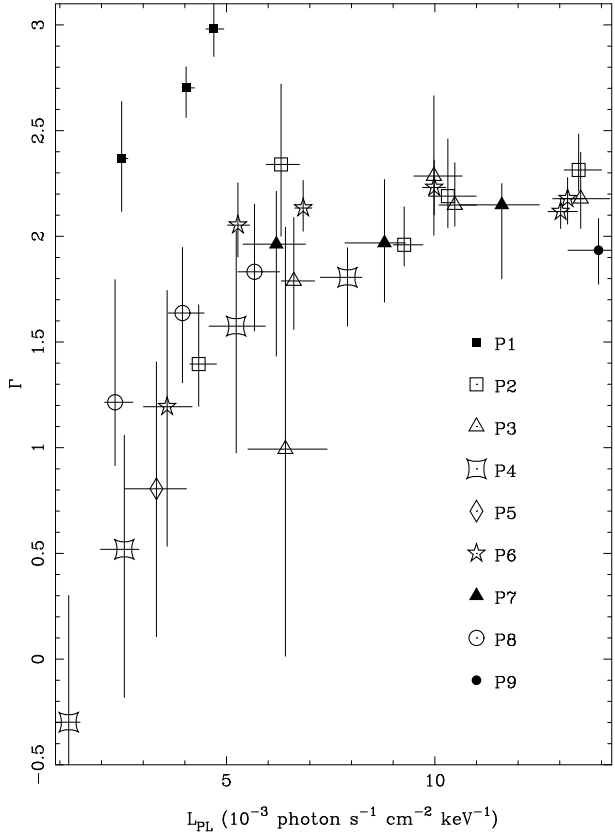,width=80mm,height=110mm}
\caption
{Behaviour of the power law component when the soft excess is
modeled with a blackbody shape.}
\label{fig:pl}
\end{center}
\end{figure}

It is obvious from the variable R4 and R7 count rates, that the
power law component is variable in amplitude. 
To investigate variability of the power law slope and the soft excess, 
we assumed a black body shape
for the soft excess (this has the lowest \(\chi^{2}\) of the three shapes
tried) and fitted the spectra with a single soft excess temperature for each
observation. The \(\chi^{2}\), fitted columns and fitted temperatures are given
in Table \ref{tab:fitting}, model B. This model is an acceptable fit to the
data at the 95\% confidence level. The models and data are shown in
Fig. \ref{fig:allspec} and each spectrum is labeled with the best fit power law slope and black body
luminosity. Confidence contours for these two
fitted parameters are shown in Fig. \ref{fig:allcont}. In each observation, a
single value of the black body luminosity can be found that is consistent with
all the different count rate spectra, \ie there is no detectable change of soft
excess flux within any of the observations. However, there is a trend for
harder power laws at lower R4 count rates. This trend is more obvious 
when the power law slope and normalization are plotted in
Fig. \ref{fig:pl}. The trend for harder slopes at low normalization is seen
when the different observations are compared {\it as well as} in
the individual observations. The three outlying points to the top left of
Fig. \ref{fig:pl} are all from observation P1, and the difference between the
power law parameters from this observation and the others is likely due to the
much smaller source circle in P1 and the different gain state of the PSPC; note
that the trend for harder power laws at lower normalization is still present
within the P1 observation itself. 

%use /disk/xray/mjp/mkn/newspec/pl.qdp

\subsection{Overall Characterization of the \Mrk\ PSPC Spectrum}

The spectral fitting described above shows that:
\vspace{3mm}

\noindent
1) The PSPC spectra can be adequately described with a power law component,
a soft excess component, and a neutral absorbing column.

\noindent
2) There is no evidence that the neutral absorbing column changes within any 
observation.

\noindent
3) The soft excess component is well characterized by a black
body shape peaking at \(<\) 0.3 keV.

\noindent
4) Variability of the soft excess component is not detected.

\noindent
5) The power law component becomes harder when it becomes fainter.
\vspace{3mm}

However, points 4) and 5) in particular are somewhat model dependent in that
the spectral fitting process depends on the exact shape of the soft excess
(which is not known) and
in that the different model components are strongly correlated with each other
in the fitting.  
The three X--ray bands R1L, R4 and R7 
will be used in the next section to investigate the spectral
variability of \Mrk\  in a way that
is {\it not} strongly model dependent. 

\section{Three Colour and Hardness Ratio Variability}
\label{sec:colvar}

We now return to the lightcurve and hardness ratios shown in Section 
\ref{sec:lightcurves}.
In light of the spectral modeling results presented in 
the previous section, 
the variability in the three X--ray bands R1L, R4 and R7 
is examined and related to variability 
of the power law and soft excess components.                
Each observation will be examined separately to ensure 
that the results are robust against the different positions that 
\Mrk\  has on the PSPC in different observations, and to ensure 
that the temporal/spatial uncertainty in the PSPC gain has 
a minimal effect. Variability in observations P5 and P9 can not 
be examined in detail because they were only a single \ros\ 
orbit (5760 seconds) in duration.

% made using /disk/xray/mjp/mkn/keep-results/allcol.pro, using allcol.com
% now made using /disk/xray/mjp/mkn/newcounts/allcol.pro using idl < allcol.dat
\begin{figure*}
\begin{center}
\leavevmode
\psfig{file=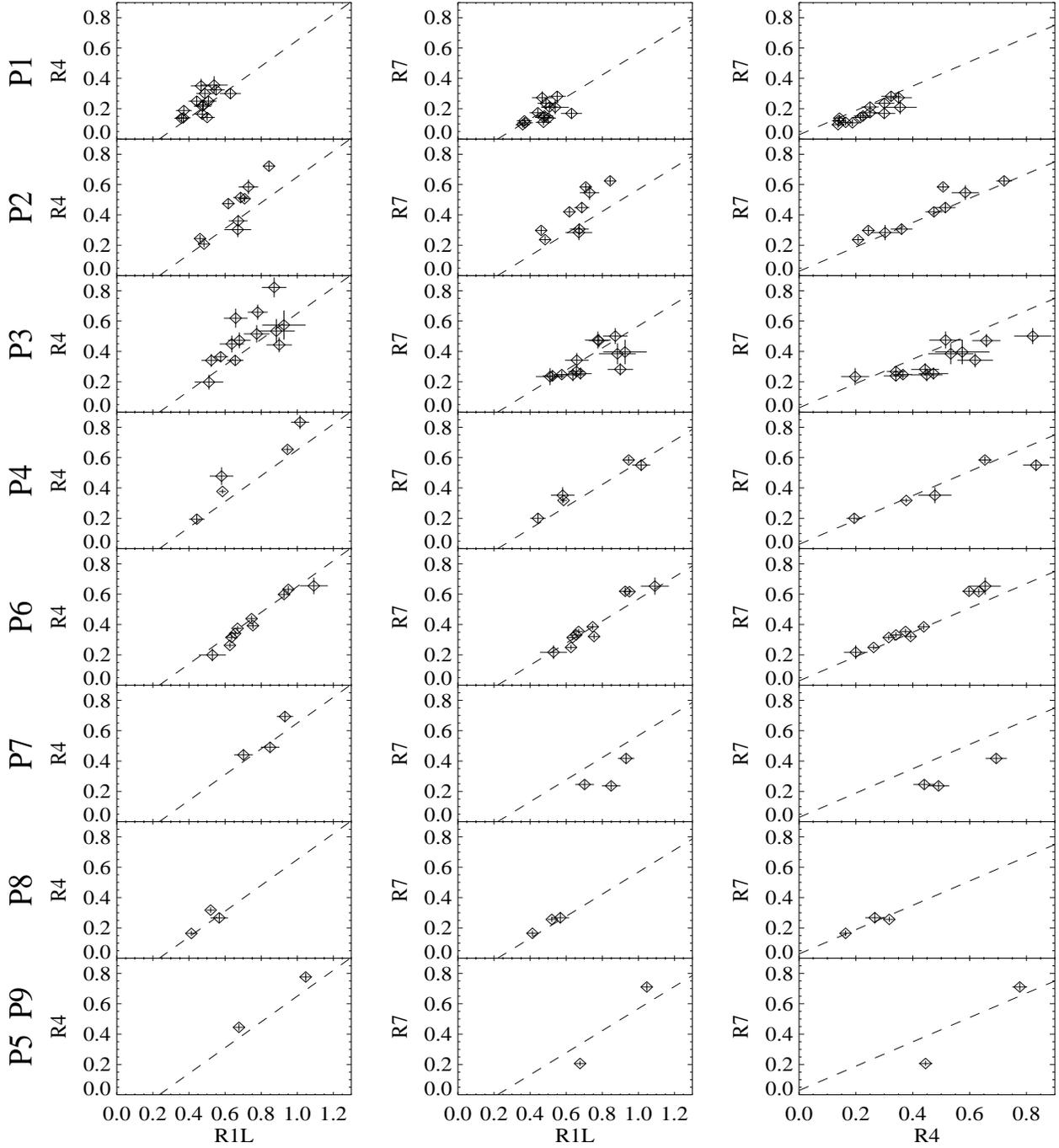,height=180mm,width=165mm}
\caption
{Comparison of the count rates for \Mrk\  in the 3
X--ray bands R1L, R4 and R7. Each observation is shown separately, except P5
and P9 which are shown together. The same dashed lines are
plotted for each observation, and represent linear fits to
the combined 3 colour data from all the observations.}
\label{fig:3countrates}
\end{center}
\end{figure*}

The count rates in the three X--ray bands are strongly correlated with each
other, both between and within the observations, as shown in Fig.
\ref{fig:3countrates}. 
When data from all the observations are
combined, the probability that any of the pairs of bands are not linearly 
correlated is
\(<10^{-18}\); 
the dashed lines in Fig. \ref{fig:3countrates} are the best fitting straight 
lines to the (R4,R1L), (R7,R1L), and (R7,R4)
relations, and comparison with data from individual observations reveals that
the source always varies in a similar fashion.
Treating the dashed lines as crude estimates of the (R4,R1L), (R7,R1L), 
and (R7,R4) 
relations, it is seen that when R7=0 (\ie the power law component has
disappeared) there is still considerable flux
(\(\sim 0.2\) counts/s) in the R1L band. The same is true when R4=0, but the
dashed line for (R7,R4) passes closer to (0,0). 
This `residual' flux in R1L betrays the presence of the soft excess, and
confirms that it peaks in the R1L band.

\subsection{Changes in the Spectrum}
\label{sec:specchange}

The variability can be examined in more detail by
comparing hardness ratios with count rates in the three bands, and allows
questions to be asked such as  
whether the change in power law flux is accompanied by a
change in power law slope, and whether variability of the soft excess, 
as well as variability of the power law, is required by the data.
The most robust comparisons are between \(HR_{hard}\) and R1L, and between 
\(HR_{soft}\) and R7. R1L is not used in the calculation of
\(HR_{hard}\), hence  
the errors on the two quantities are independent; similarly for R7 and
\(HR_{soft}\). 
We will examine changes in hardness ratios within the five longest
observations: P1, P2, P3, P4 and P6; the other observations have too few
data points to examine hardness ratio changes properly.

These hardness ratios will be compared to models of how they should behave 
as the different components vary. We will describe the soft excess using the
parameters appropriate for a black body absorbed by a neutral column of  
\(3 \times 10^{20}{\rm cm^{-2}}\).
 However,
the detailed shape of the soft excess 
does not matter for the three colour study because the soft excess 
is only observed in the two bands R1L and R4: any change in shape of the soft
excess has the same effect as changing the black 
body temperature, \ie it changes the contribution of the soft excess to 
R4 relative to R1L.

From the spectral modeling performed in Section \ref{sec:spectral}, 
one can estimate the extent of
parameter space which requires investigation. 
Table \ref{tab:fitting} 
shows that when the
soft excess is described as a blackbody, most data are consistent with
a temperature \(T\) 
between 
60 eV and 80 eV. The maximum fitted temperature of 96 eV, and 
a blackbody
component at this temperature or less has an R7 count rate which is at least 45
times smaller than the R1L and R4 count rates, even when 
attenuated by a neutral absorbing column of 
\(3 \times 10^{20}{\rm cm^{-2}}\). Similarly, when the soft 
excess is modeled as two black bodies, the contribution to R7                
is negligible.
Even when the soft excess is modeled using the much broader bremsstrahlung
shape, the highest fitted bremsstrahlung temperature (\(\sim\)0.2 keV) leads 
to an
R7 count rate which is 15 times smaller than the R1L or R4 count rates.
We can therefore assume 
that the soft excess, whatever its shape, 
provides a negligible contribution to the R7 count rate.
%The mean fitted blackbody temperature of \(\sim 70\) eV
% (see model B, Table \ref{tab:fitting}),
%absorbed by 
%\(3 \times 10^{20}{\rm cm^{-2}}\) of neutral material, produces an R1L/R4
%ratio of 4/1. 
%Therefore, we can reasonably assume that the soft excess will provide
%the majority of its contribution in the R1L band; the power law component
%must dominate the R7 band, and probably dominates the R4 band as well.

Fig. \ref{fig:allcont} shows that the black body soft excess
normalization \(L_{BB}\) lies between 3 and 9 
(\(\times 10^{37} {\rm erg\ s^{-1}}/D^{2}\), where \(D\) is the distance to
\Mrk\  in kpc), and the majority of fitted power law slopes have 
\(1.7 < \Gamma <2.3\).

\begin{figure*}
\begin{center}
\leavevmode
\psfig{file=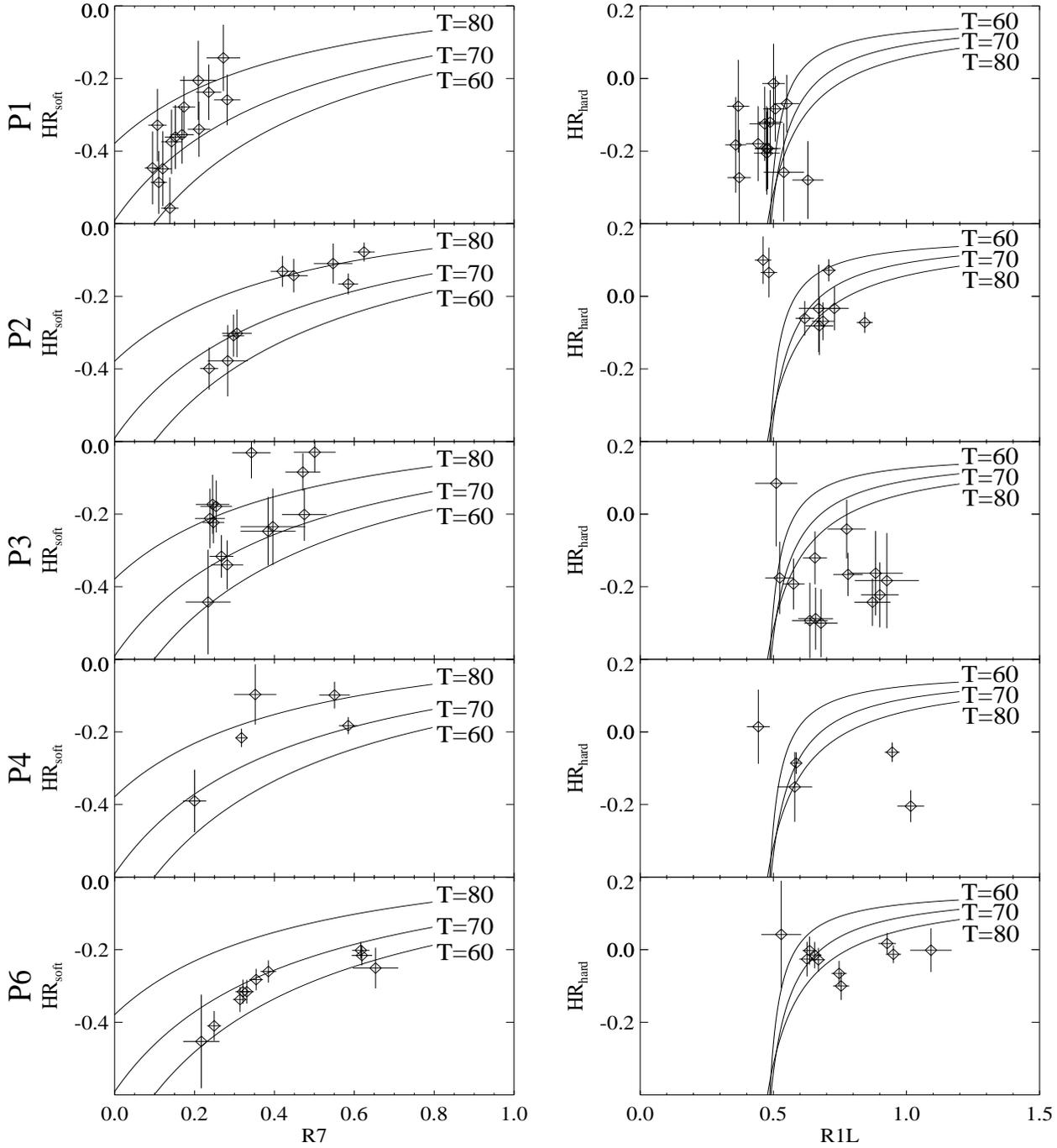,width=165mm,height=180mm}
\caption{$HR_{soft}$ as a function of (hard)
R7 count rate and $HR_{hard}$ as
a function of (soft) R1L count rate. The model curves are for flux
variability of a $\alpha=1$ power law; the soft excess has normalization
$L_{BB}=
6 \times 10^{37} {\rm erg s^{-1}}/D^{2}$ (where $D$ is the distance in kpc),
and temperature $T$ in eV as indicated on the curves.}
\label{fig:varitemp}
\end{center}
\end{figure*}

\begin{figure*}
\begin{center}
\leavevmode
\psfig{file=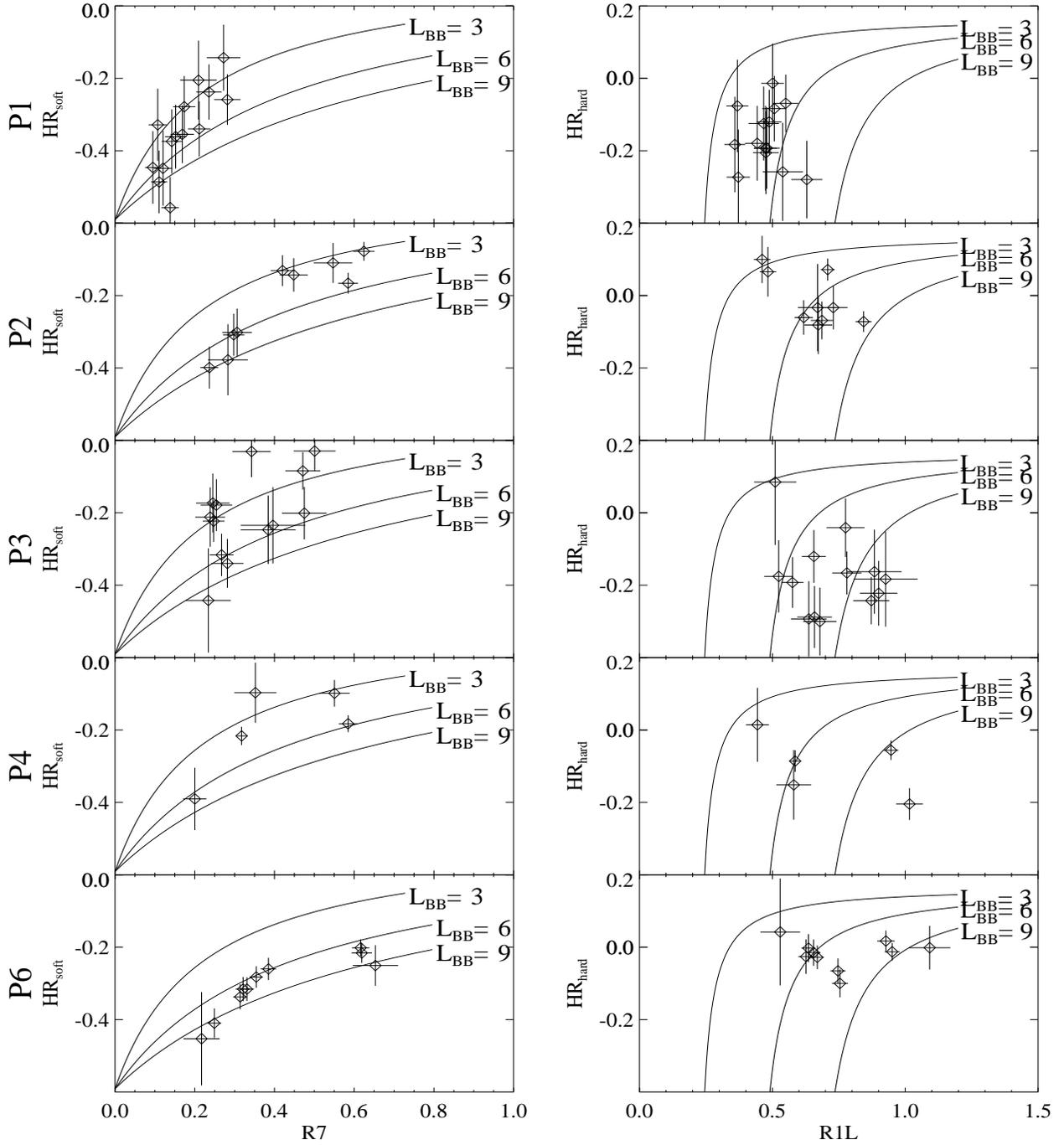,width=165mm,height=180mm}
\caption{$HR_{soft}$ as a function of (hard)
R7 count rate and $HR_{hard}$ as
a function of (soft) R1L count rate. The model curves are for flux
variability of a $\alpha=1$ power law; the soft excess has a temperature of
70 eV and
normalization
$L_{BB}$ 
(in units of $10^{37} {\rm erg s}^{-1}/D^{2}$ where $D$ is the distance in kpc),
as indicated on the curves.}
\label{fig:varilum}
\end{center}
\end{figure*}

\begin{figure*}
\begin{center}
\leavevmode
\psfig{file=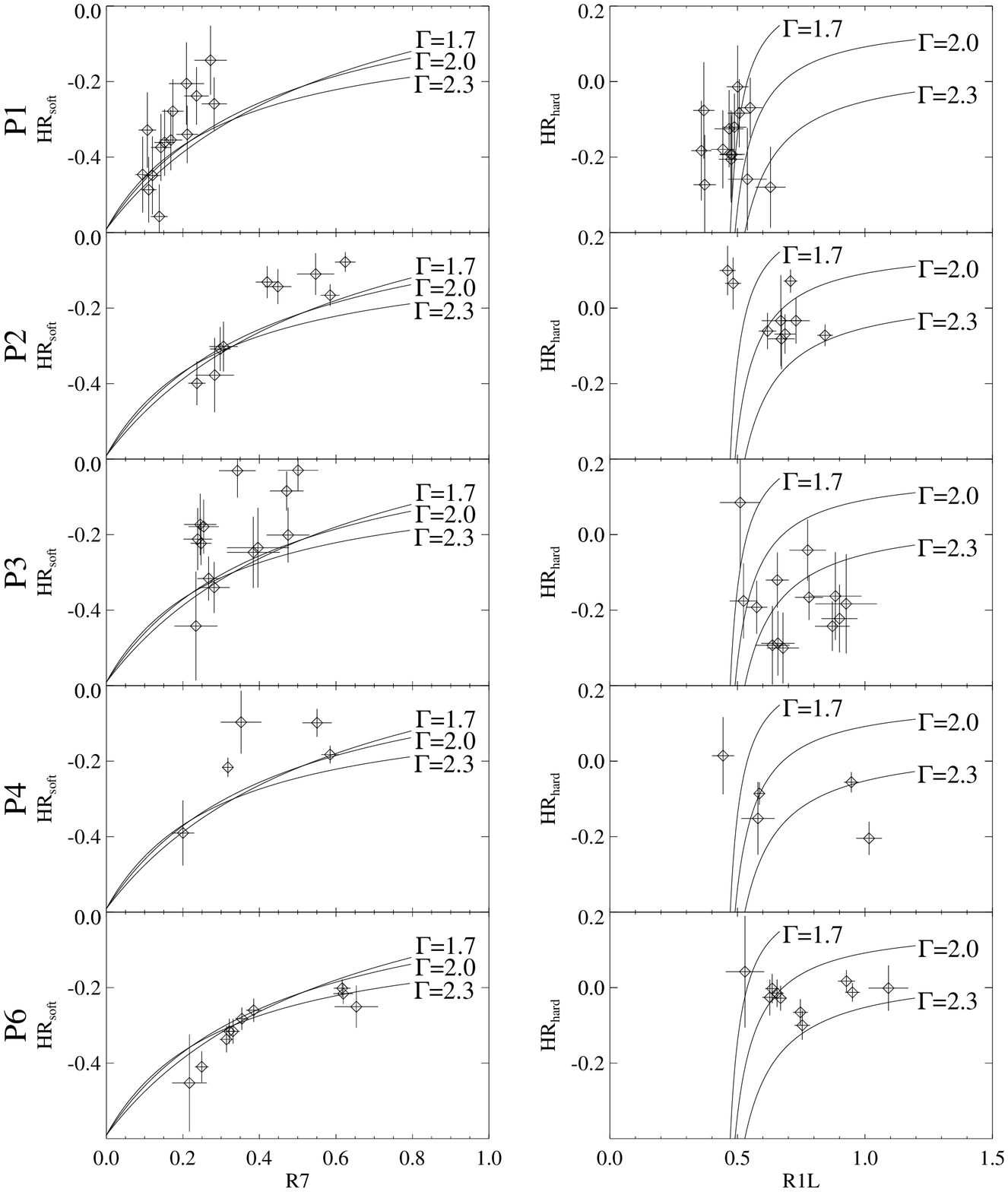,width=165mm,height=180mm}
\caption{$HR_{soft}$ as a function of (hard)
R7 count rate and $HR_{hard}$ as
a function of (soft) R1L count rate. The model curves are for flux
variability of power laws with slopes $\alpha=0.7$, $\alpha=1.0$ and
$\alpha=1.3$.
The soft excess has temperature $T=70$ eV and normalization
$L_{BB}=
6 \times 10^{37} erg s^{-1}/D^{2}$ (where $D$ is the distance in kpc).}
\label{fig:varislope}
\end{center}
\end{figure*}

The hardness ratios are plotted in Fig. \ref{fig:varitemp} and repeated in 
Figs. \ref{fig:varilum} and \ref{fig:varislope}. 
Each solid curve is
 the expected locus of hardness ratios for pure flux variability (\ie no
change in slope) of the
power law component, with a constant soft excess; curves are
provided for a range of soft excess
temperatures in Fig. \ref{fig:varitemp}, soft excess normalizations in Fig.
\ref{fig:varilum} 
and power law slopes in Fig. \ref{fig:varislope}.
Note that we have made no attempt to `fit' the data in this section.

%First find values for the blackbody component.
%When power law is gone,maximum BBT is 70, max BBL is 4
%With maximum(temp,normalisation)BB,no single power law can describe everything.
%made these plots with /disk/xray/mjp/mkn/keep-results/varislope.com,
%varilum.com, and varitemp.com

The relation between \(HR_{soft}\) and the R7 count rate can be reproduced
well by pure flux variability of the power law. For any combination of soft
excess parameters and power law slopes, power law
flux variability predicts that
\(HR_{soft}\) should increase with R7,  which is the observed behaviour.
 
However, pure flux variability of the power law predicts that \(HR_{hard}\)
is a monotonically increasing function of R1L count rate. This does not match
the observed behaviour of \(HR_{hard}\), which {\it does not} increase with
R1L for
R1L \(<0.8\) as seen in Figs \ref{fig:varitemp} to \ref{fig:varislope}. 
The relation between
\(HR_{hard}\) and R1L can be reproduced by a steepening of the power law
with increasing power law flux (see Fig. \ref{fig:varislope}) and/or
an increase in the soft excess
normalization (see Fig. \ref{fig:varilum}) with increasing power law flux.
In the latter case, to reproduce the variability of \(HR_{hard}\), the soft 
excess must vary by a factor of \(\sim 3\), and
since the higher soft excess normalization is required at higher R4
count rates its variability must be correlated with that of the power law
component.

\subsection{Variability Amplitude}

A simple means of examining the variability of the different spectral
components is to compare the variability amplitude in the three X-ray bands.
For this we have used the normalized variability amplitude, which is
defined as the standard deviation \(\sigma_{int}\)
divided by the mean count rate. This is
corrected for measurement errors by assuming that
\begin{equation}
\sigma_{int}^{2}=\sigma_{obs}^{2}-\sigma_{err}^{2}
\label{eq:sigma}
\end{equation}
where \(\sigma_{int}\) is the true standard deviation, \(\sigma_{obs}\) is
the observed standard deviation,
and \(\sigma_{err}\) is the measurement error (see Edelson 1992).

\label{sec:variamp}

\begin{table*}
\caption[Normalized variability amplitude in the three X--ray bands]
{Normalized variability amplitude in the three X--ray bands
for each of the observations.}
\label{tab:variamp}
\begin{tabular}{ccccc}
&&&&\\
\ros\ & number & R1L variability & R4 variability & R7 variability \\
pointing & of points & amplitude & amplitude & amplitude \\
&&(\%)&(\%)&(\%)\\
\hline
&&&&\\
P1&14&12\(^{+4}_{-2}\)&32\(^{+10}_{-6}\)&31\(^{+10}_{-6}\)\\
P2&9&19\(^{+9}_{-4}\)&34\(^{+16}_{-7}\)&33\(^{+16}_{-7}\)\\
P3&13&17\(^{+6}_{-3}\)&29\(^{+10}_{-5}\)&29\(^{+10}_{-5}\)\\
P4&5&35\(^{+19}_{-12}\)&48\(^{+25}_{-16}\)&41\(^{+22}_{-14}\)\\
P6&10&20\(^{+9}_{-4}\)&38\(^{+16}_{-8}\)&40\(^{+17}_{-8}\)\\
P7&3&13\(^{+23}_{-4}\)&22\(^{+40}_{-7}\)&35\(^{+63}_{-11}\)\\
P8&3&9\(^{+16}_{-3}\)&32\(^{+58}_{-10}\)&19\(^{+34}_{-6}\)\\
\end{tabular}
\end{table*}

The variability amplitude in the three bands is shown in Table
\ref{tab:variamp}. The 1\(\sigma\) error quoted is calculated from equation 2 in Done
\etal (1992a) and is the uncertainty in the width of an assumed underlying 
Gaussian distribution of count rates from which the data were drawn; note that
this is not the measurement error of the count rate,
which has already been subtracted in quadrature (Eq. \ref{eq:sigma}).
In {\it
every}
observation the variability is
smaller in the R1L band than in either R4 or R7. This shows that the more
variable component is the
power law, which dominates the flux in R4 and R7 bands.
The reduced
variability amplitude in R1L shows that the soft excess is relatively
invariant, 
diluting the overall (power law + soft excess) R1L variability.
This means that to explain the hardness ratio variability in Section
\ref{sec:specchange}, the power law slope {\it must} vary.
% note that the soft excess must change by a substantial amount, correlated
% with the power law if the power law does not change slope, and this is not
% supported by the variability amplitude.

\section{Cross Correlation Analysis}
\label{sec:crosscorrelation}

%If the soft excess and power law components are emitted from
%physically distinct regions near the centre of \Mrk, then some time delay is
%expected between changes in the two components, and can be used to
%distinguish between different models for their production (see Section
%\ref{sec:models}). 

The time dependence of flux changes at different energies is 
potentially a 
powerful discriminator between different models for the emission mechanisms 
(see Section \ref{sec:models}).
It has been studied by cross correlating the 
count rates in the three bands using the discrete
correlation function (DCF hereafter, see Edelson and Krolik 1988).

The DCF for the R1L and R4 lightcurves is defined as
\small
\[{\rm DCF}(\tau)=\frac{1}{M}{\large \sum} 
\frac{(R1L(t)- \langle R1L \rangle)(R4(t+\tau)- \langle R4 \rangle)}
{\sigma_{R1L}\ \ \ \sigma_{R4}}\]
\normalsize
where the sum is made over all M values of t for which there is a
measurement of R1L at time t and a measurement of R4 at time t+\(\tau\).
 \( \langle R1L \rangle \) and \( \langle R4 \rangle \) are the mean 
R1L and R4 count rates respectively;
\(\sigma_{R1L}\) and \(\sigma_{R4}\) are the standard deviations of the R1L
and R4 count rates respectively and have been corrected for measurement error
using equation \ref{eq:sigma} in Section \ref{sec:variamp}.
The standard error on the DCF is defined as 
\small
\begin{equation}
\sigma_{DCF}(\tau)=\frac{1}{M-1}\times
\label{eq:dcferror}
\end{equation}
\[\sqrt{\sum 
\left[\frac{(R1L(t)- \langle R1L \rangle )(R4(t+\tau)- \langle R4 \rangle )}
{\sigma_{R1L}\ \ \sigma_{R4}}-DCF(\tau) \right]^{2}}\]
\normalsize
The DCF is similarly defined for any combination of R1L, R4 and R7.
This method is appropriate
to our \ros\ data since it requires no
interpolation, and provides simple error estimates.  

\ros\ wobbles with a period of 400s, preventing the wire grid of the PSPC
window from permanently masking sources; 400s is the smallest time binning
which does not introduce spurious variability associated with the spacecraft
wobble.
Hence lightcurves with
400s time bins have been cross correlated. 
Due to 
the orbital interruptions in \ros\
observations the number of pairs available for cross correlation is very
small for time differences greater than 1200 seconds.
Only the five longest observations have been used: the others are 
too short to provide a useful
number of pairs for cross correlation. 

%used table in disk$a1:[mjp.mkn766.timing]

\begin{figure*}
\begin{center}
\leavevmode
\psfig{file=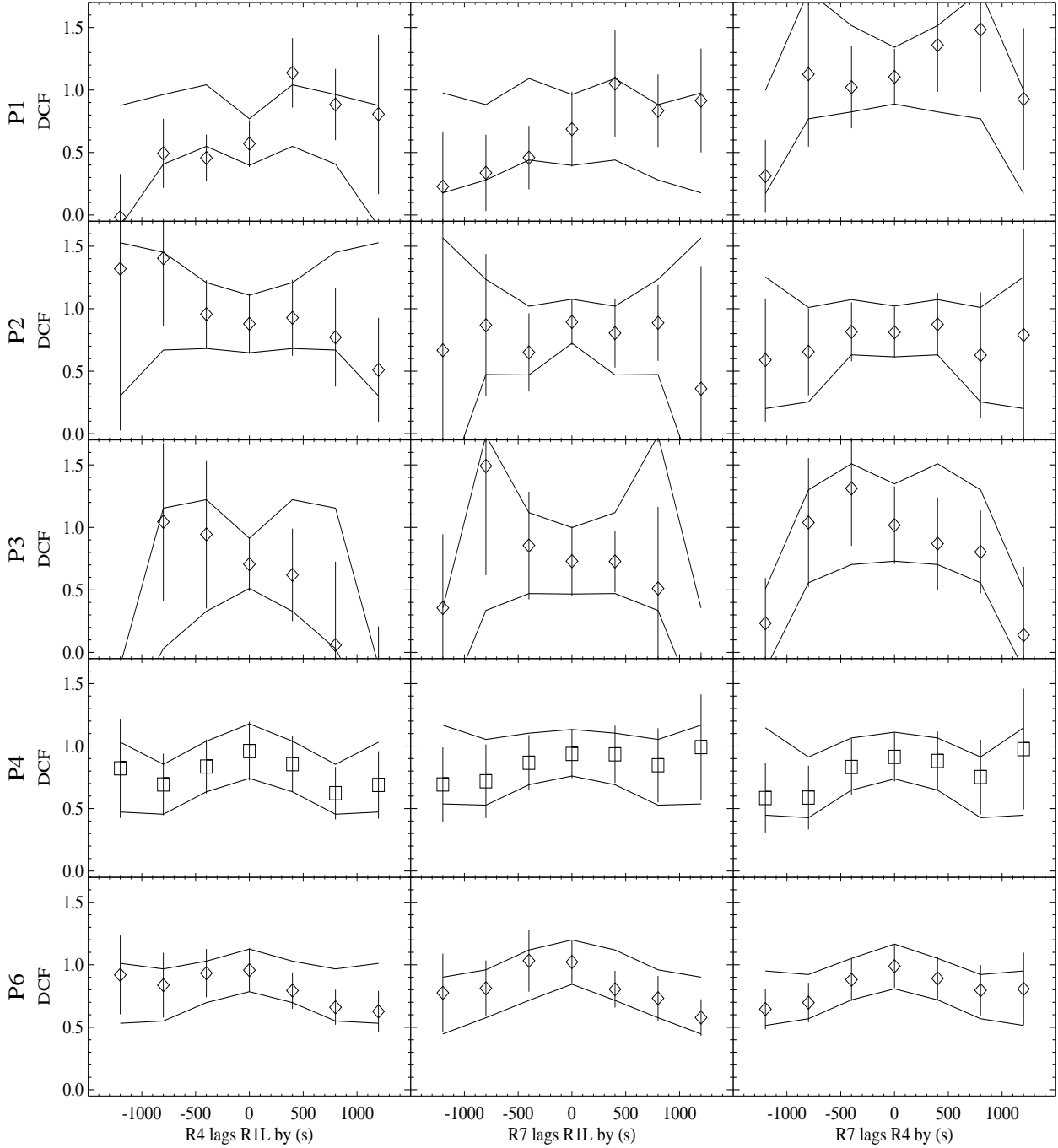,width=165mm,height=180mm}
\caption
{Discrete correlation function (DCF) of R1L, R4, and R7 for the five
long observations. The solid lines are the bootstrap 
68\% confidence limits to the DCF
if it is symmetric.}
\label{fig:ccor}
\end{center}
\end{figure*}

The results of the cross correlation analysis are shown in Fig. \ref{fig:ccor}.
The
error bars are 68\% as given by equation \ref{eq:dcferror} above. The cross
correlation was intended to reveal whether changes at the soft energies
precede changes in the hard power law or vice-versa. The time
bins are the same for all three bands, so there is no asymmetry in the DCF
originating from the temporal sampling. Since the
points of the DCF are not independent, we require some simple
means of testing the significance of the observed DCF against the null
hypothesis that neither
component leads the other. We have used a bootstrap error
method: for each time lag, pairs of points were chosen at random from all the
original pairs with corresponding positive {\it or} negative time lag
and the DCF constructed with the same number of pairs as the real DCF; this
was repeated 1000 times and the solid line in Fig. \ref{fig:ccor} 
corresponds to the
region bounding 68\% of the symmetrical bootstrap simulations.
It is seen in Fig. \ref{fig:ccor} that the only significant deviation from the
symmetrical bootstrap simulations is 
in observation P1 where there is a marked
asymmetry in the DCF of R1L and R4. To improve the signal
to noise further, the DCF has been computed using a single bin for positive 
time lags and a
single bin for negative time lags. Applying
our bootstrap error method we find that 
the observed asymmetry of the DCF of R1L and R4 is just significant at the 
95\% 
level,
and that of R1L and R7 is significant at 85\% for 
observation P1. The DCF is within the 68\% confidence limits for R7
and R4 in observation P1 and for all combinations of R1L, R4 and R7
for the other observations. 

Thus, in P1, when \Mrk\  is at its faintest, and consequently the power law
component is smallest relative to the soft excess, changes at \(< 0.5\) keV  
appear to precede changes at \(> 0.5\) keV.
However, it must be pointed out that the cross correlation has been performed
on 5 datasets, of which only one has 95\% significant asymmetry; 
the probability of one or more dataset having this level of asymmetry, by
statistical fluctuation alone, is 23\%. 

\section{Discussion}
\label{sec:discussion}

\subsection{Long and Short Term Variability Properties of \Mrk}

There is no apparent long 
term variability trend present in the \ros\ observations
of \Mrk; changes in the \ros\ count rate between observations are no larger
than the changes seen in individual observations. 
On a time scale of hours, \Mrk\  varies continuously in every
observation period. The spectral variability takes the same
form in every observation, 
which can be explained by variability of the power law
slope and normalization.

Given that the behaviour is the same throughout the PSPC
pointed observations from 1991 to 1994, 
the same spectral variability would be expected in the \AS
observation of \Mrk\  in 1993.
Leighly \etal (1996) report that the major change in the \AS band was that
the power law slope became steeper as the power
law flux in the \AS band increased.  
This behaviour is entirely consistent
with the spectral variability seen in the \ros\ observations and reported here.
Similarly, in the \exo ME observations of 1985 and 1986
steeper power law photon indices are found for increased power law
normalization (Molendi, Maccacaro \& Schaeidt 1993).

\subsection{The Origin of the Power Law}

There are two popular 
models for the power law X--ray component of radio quiet AGN; both are based
on Compton upscattering of UV or soft X--ray photons from an accretion disk
or  optically thin plasma. In the first model,
the observed power law emission is simply that from upscattering of the 
soft photons in 
a hot (kT \(\sim\) 500 keV) plasma, or by
a non - thermal distribution of relativistic electrons (Walter \& Courvoisier
1992).
In the second model, the emission originates in a region in which the radiation
density is large enough that the region is opaque to photon - photon
collisions (\(\gamma\)--ray - \(\gamma\)--ray 
or \(\gamma\)--ray - X--ray) 
which give rise to electron - positron pairs. These pairs
down scatter the  \(\gamma\)--ray radiation 
and produce the observed X--ray spectrum,
including the soft X--ray excess
(Zdziarski \etal 1990).
{\it Both} these models predict that in the variability of a single
source the power law slope is softer when the
power law flux is higher (Done \& Fabian 1989, G. Torricelli-Ciamponi \&
Courvoisier 1995) as observed in \Mrk. 

Leighly \etal (1996) show that \Mrk\  has sufficient radiation density in
the emission region that pair
production may be important. The compactness parameter is defined as 
\[l=L\sigma_{T}/Rm_{e}c^{3}\]
where \(L\) is the luminosity, \(R\) is the size of the source,
\(\sigma_{T}\) is the Thompson scattering cross section, \(m_{e}\) is the
mass of the electron and \(c\) is the speed of light; pair production may be
important if \(l>10\) (Svensson 1987). 
Using the shortest factor 2 variability time from 
the \AS lightcurve (\(\Delta t \sim 1000s\)) to define the
size of the source, assuming \(R <c \Delta t\), Leighly \etal found
\(l \sim 12\). 
Leighly \etal (1996) propose 
that the variability observed during the \AS observation is 
best explained by pair reprocessing, because the major
spectral variability was confined to a single rapid increase in the flux of
\Mrk\ (after which it remained relatively constant), and because the soft excess
is not required to fit the high count rate spectrum. This would be expected
if the energies of the relativistic electrons were to increase suddenly: the
first order pair reprocessed spectrum (seen as the soft excess) is replaced
by a non-linear pair cascade, 
which produces the \(\alpha \sim 1\) power law spectrum.

However, from the \ros\ observations it is clear that the soft excess {\it
does not} disappear when the power law increases in flux and slope, as this
would lead to exactly the opposite behaviour of \(HR_{hard}\) from that seen
in figs \ref{fig:varitemp} to \ref{fig:varislope}.

If the spectral changes in a source of compactness \(l \sim 12\) 
are determined by pair reprocessing, the hard X--rays would be expected to
vary \(\sim \) half a light crossing time before the soft X--rays (Done \&
Fabian 1989). Assuming
1000s as the light crossing time for \Mrk\  this delay is \(\sim\) 500s. This
should lead to a significant asymmetry in the cross correlations in Fig.
\ref{fig:ccor}
such that changes in the R7 band (and to a lesser extent R4 band) 
lead the changes in the R1L band. No such asymmetry is seen (the DCF from
observation P1 is
asymmetric in the opposite sense).

If the power law component is produced by 
relativistic particles
scattering soft photons and
pairs are not important, then changes in the power law component 
may be simultaneous at all energies.
If the variability of the power law component is driven by changes in the
number of soft seed photons (\ie variability of the soft excess) then this
could explain the apparent asymmetry, with soft flux leading the hard flux, 
in the cross correlation of observation P1. However, 
in current models (\eg Torricelli-Ciamponi \& Courvoisier 1995), 
the change in spectral index with power law flux is
brought about by a change in the relativistic electron energy spectrum, not
a change in the number of soft photons, and hence there is no expectation
that the soft flux should change first. This would explain how the 
power law varies when there is no detectable soft excess variability, and 
is consistent with the 
behaviour of 
\(HR_{\rm hard}\) in all the observations.

\subsection{The Origin of the Soft Excess}

The low variability amplitude in the R1L band compared to the harder R4 and
R7 bands (see Table \ref{tab:variamp}) 
would suggest that the soft excess is relatively
invariant on short time scales compared to the power law component.
This suggests that the soft excess and power law 
are different components, and hence a pair reprocessing origin for the
soft excess is not likely.

The relative invariance of the soft excess  also rules out models whereby the
soft excess is power law emission reflected or reprocessed 
close to the central regions. However, 
if the soft excess were power law emission 
reprocessed at regions some distance (light days or more) 
from the central regions,
then the soft excess variability would be smeared out by the long response;
such a model is therefore compatible with the \ros\ data.

The lack of short term soft excess variability  
is consistent with the it being
part of the big blue bump,  
which is known to have a longer variability
time scale than the power law X--ray emission in radio quiet AGN 
(\eg NGC4051, Done \etal 1990). 
The soft excess spectrum is well fit by one or two black bodies, and hence is 
consistent with the shape expected from the high energy tail of a hot
accretion disk (\eg Ross, Fabian \& Mineshige  1992.)

It is not possible to determine if the soft excess of \Mrk\ varies over a 
longer
time scale than an observation;
different values for the blackbody parameters are
obtained from different observations, but the difference in off-axis angles
between the observations, and the gain drift of the \ros\ PSPC, mean
that long term soft excess variability cannot be assessed with any certainty. 

We note that the results presented here, \ie that short term
variability in the soft excess component is not detected, 
are in marked contrast with the
findings of Molendi \& Maccacaro (1994). Their study of \Mrk\  is based on
observation P2, in which they find that the soft excess varies with 
the power law component by approximately the same amplitude, and hence that
the two components are related to the same physical process or are causally
connected. However, they determine the blackbody and power law parameters
by fitting models to spectra in four states defined by overall count rate. 
They do not quote uncertainties on the power law and
blackbody 
normalizations; the apparent correlation between blackbody  and power law 
normalization may be caused by the coupled nature of the blackbody
temperature and power law slope when fitting low resolution \ros\ spectra. 
The three colour data (see Fig. \ref{fig:3countrates}), 
and lower variability amplitude in  R1L than
the R4 and R7 (Table \ref{tab:variamp}), for
all the \ros\ observations, including P2, show that the blackbody
flux {\it can not} vary as much as the power law flux.
Consequently, the model presented by Molendi \& Maccacaro whereby 
changes in the accretion rate of an accretion disk determine the variability
time scales in the \ros\ band is not valid, 
because the soft excess, assumed to be the
high energy tail of an accretion disk, probably does not vary with the 
time scales found in the \ros\ lightcurve.

\section{Analogy with Galactic Black Hole Candidates}

Galactic black hole candidates (GBHCs) have
X--ray spectra that are qualitatively similar (power law component
with a soft excess, an Fe line and reflection hump), to those of Seyfert 1
galaxies. Both types of objects are thought to be powered by accretion onto
a black hole, and it is possible that the physical processes which produce
X--ray emission are the same in
both types of objects. GBHCs have two states: in
the `low' state they have  hard (\(0.3<\alpha<0.8\)) power law
spectra with low energy (\(<2\) keV) soft excesses and in the `high' state they
have steep power law components (\(1.3<\alpha<1.8\)) and powerful
soft excesses which dominate the spectra below \(\sim\) 8 keV (Ebisawa \etal
1996). In the low state, GBHCs exhibit very rapid (millisecond) variability
of the power law component, but vary less rapidly 
when they are in the high state.

The connection between GBHCs and Seyfert 1 galaxies 
has been strengthened by
the discovery that RE J1034+396 has an extremely powerful soft excess, 
a steeper power law component 
than most Seyfert 1s, and does not vary rapidly,
\ie it is like a GBHC in a high state (Pounds, Done \& Osborne 1995).
In this analogy, \Mrk\  
is equivalent to a low state GBHC in that it
shows continuous variability of the power law component, the spectral index
of which is similar to those found in low state GBHCs. The variability
time scales of GBHCs are much smaller than those of Seyfert galaxies like
\Mrk, but this is not surprising given the large difference in the central
black hole masses and hence overall dimensions inferred for
Seyfert nuclei and GBHCs.
It is believed that the soft excess in the GBHC Cygnus X-1 comes from a
region of larger spatial extent (and hence varies on a longer time scale) 
than that producing the power law emission (Done \etal 1992b); the results presented here
show that the same is probably true for \Mrk. Furthermore, changes in the soft
X-ray flux in Cygnus X-1 lead changes in the hard X--ray flux (Miyamoto \&
Kitamoto 1989), which may
 be true also for \Mrk\  in observation P1 (see Section
\ref{sec:crosscorrelation}).

Note that ultrasoft
spectra (\ros\ \(\Gamma > 3\)) are not found in Seyfert 1s with broad lines 
(FWHM \(>\) 3000 km s\(^{-1}\), Boller, Brandt and Fink 1996), 
\ie the broad line Seyfert 1s are only analogous to GBHCs in the low state. 
However, {\it both} GBHC states are represented among the NLS1s.

\section{\Mrk\  and Other NLS1 Galaxies}
\label{sec:others}
\subsection{X--ray Emission}

\Mrk\  has one of the hardest \ros\ spectra of the NLS1s
in the sample of Boller, Brandt and Fink (1996).
Indeed, some NLS1 galaxies have X--ray spectra so soft
that they are
quite unlike that of \Mrk\  (\eg WPVS 007, Grupe \etal 1995, RE J1034+396,
Puchnarewicz \etal 1995). 
It is therefore important to question whether
the X--ray emission observed in the ultrasoft NLS1s could come
from the same physical processes as that observed in the very much harder
NLS1s like \Mrk. Variability and spectral 
properties are important
diagnostics; similar physical processes can be expected to produce similar
spectral shapes and similar variability characteristics.
The results presented here show that the X--ray spectrum of \Mrk\  has two
important X--ray emission components (the power law and soft excess) which
have
different spectral shapes and different variability properties: a power law
component which varies rapidly (thousands of seconds) 
and is softer when it is  brighter, and a soft
excess which does not exhibit any measurable rapid variability.

NLS1 galaxies which do not have extremely
soft \ros\ spectra might be expected to be most similar to \Mrk. The
best studied of these at X--ray wavelengths are NGC4051 and MCG-6-30-15.
In both these AGN the power law component shows rapid (hundreds of seconds) 
variability, and
becomes softer as it increases in flux. This variability is found in 
\exo and \ginga
observations (Papadakis \& Lawrence 1995, Matsuoka \etal 1990, Kunieda \etal
1992, Pounds, Turner \& Warwick 1986), \ie has continued for a number of
years. This is exactly the behaviour found in \Mrk.

The ultra-soft NLS1 galaxies show a more varied
picture; some like RE J1034+396 show no significant rapid variability in
either power law or soft excess components (Pounds, Done \& Osborne 1995),
while IRAS 13224-3809 (Otani \etal 1996, Boller \etal 1993) shows
extremely rapid variability involving both soft excess and power law
components and 
PHL 1092 (Forster \& Halpern 1996) shows rapid variability of at least the soft
excess component.
Other ultra-soft objects have shown extreme variations in their soft excess
over a relatively long time scale (\eg the \ros\
count rate of RE J1237+264
varied by a factor of \(\sim\) 70 between two observations separated by one
year with no measurable spectral change, and the \ros\ count rate of WPVS 007 
varied by a factor of \(\sim 400\) in one year). 

Boller, Brandt \& Fink (1996) concluded that NLS1s in general 
show considerable rapid variability, as well as softer than average \ros\ 
spectra. It 
has often been assumed that the variability comes from the soft excess in these
objects (a natural assumption given that NLS1s as a class 
are unusual in showing both
large soft excesses and rapid soft X--ray variability) but this is not the case
for \Mrk.  

%which has implications
%for the sizes of the emission regions; if the rapid
%variability is confined to the power law component, then the maximum 
%size of the
%soft excess emission region is not given by the 
%minimum \ros\ count rate doubling
%time scale. 
%In this case, the doubling time scale for the \ros\ count rate {\it does}
%give 
%an upper
%limit to the size of the power law emission region.

\subsection{Multiwavelength Spectra}

\begin{figure}
\begin{center}
\leavevmode
\psfig{file=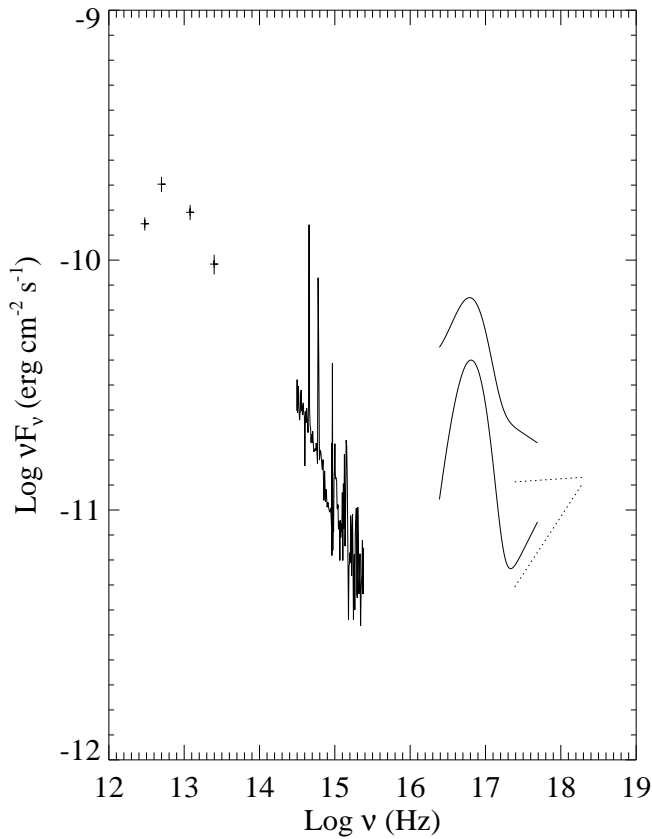,height=110mm,width=85mm}
\caption{The multiwavelength spectrum of \Mrk}
\label{fig:multi766}
\end{center}
\end{figure}

The multiwavelength spectrum of \Mrk\  is shown in Fig. \ref{fig:multi766}. 
The largest and
smallest blackbody and power law model components 
from the P6 \ros\ spectral fitting
are shown as solid lines. Also shown, as dotted lines, are the power law
components from the two spectral states identified in \AS data by Leighly
\etal (1996). The optical spectrum is taken from Gonz\'alez-Delgado \& P\'erez
(1996).
Note that the data in this plot are not simultaneous, the optical and
ultraviolet spectra have not been dereddened, and the model X--ray spectrum has
not been adjusted for absorption by neutral material.
\Mrk\ has a large \IRAS bump: it emits more power from 10 - 100
microns than in the \ros\ energy range.

\begin{figure}
\begin{center}
\leavevmode
\psfig{file=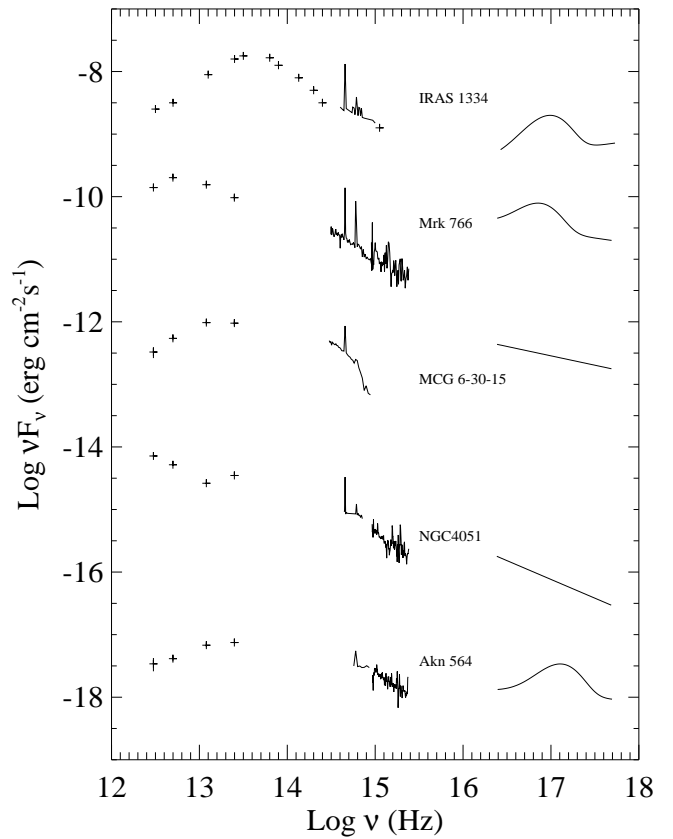,width=85mm,height=110mm}
\caption{The multiwavelength spectra of the outlying NLS1s of Walter \&
Fink (1993). Note that the spectra have been vertically shifted for clarity.}
\label{fig:multiall}
\end{center}
\end{figure}

The spectrum is unusual among Seyfert 1 galaxies in that it shows no
evidence for the big blue bump in the ultraviolet. This has been noticed
before by Walter \& Fink (1993) in that \Mrk\  has an unusually low ratio of
ultraviolet to soft X--ray flux given its soft X--ray slope. Most of the 
Walter \& Fink (1993) sample of
\ros\ all sky survey selected Seyfert 1 galaxies 
fit a strong correlation between soft X--ray slope and the ratio of 
ultraviolet (1375 \AA) to soft X--ray flux, suggesting that they have big
blue bumps which  rise in the UV, peak in the EUV and extend to soft X--ray
energies. 
Besides \Mrk, five other outliers were found by Walter \& Fink (see
their figure 8),
which have a low ultraviolet to soft X--ray flux ratio given their soft
X--ray spectral  slopes. Of these five, one is the broad line Seyfert 1
galaxy IC4329A, in which the lack of ultraviolet flux is probably caused by
attenuation by dust in the edge on host galaxy.
The remaining four Seyferts, Akn 564, NGC
4051, IRAS 13349+2438, and MCG-6-30-15 
have Balmer lines with FWHM \(<2000\) km s\(^{-1}\) and 
could all be classified as NLS1s by the criteria of Goodrich
(1989). The multiwavelength spectra of these objects are plotted in Fig.
\ref{fig:multiall}, and sources for these data are given in Table
\ref{tab:multiall}.
All are particularly similar to that of  \Mrk; they have very large
infrared bumps and falling optical -- ultraviolet spectral shapes.
The lack of ultraviolet flux in these
objects may be a result of their having very high temperature soft excesses,
or (as suggested by Walter \& Fink 1993) 
it may be due to reddening by dust. Although not included in the Walter \& 
Fink sample, the ultrasoft NLS1 RE J1034+396 also exhibits a similar 
multiwavelength continuum shape (Puchnarewicz \etal 1995).

\begin{table*}
\caption{Data sources for Fig. \ref{fig:multiall}}
\label{tab:multiall}
%\scriptsize
\begin{tabular}{@{\hspace{0mm}}lcccc@{\hspace{0mm}}}
&&&&\\
Object&$\nu F_{\nu}$ shift&IR&Optical&X--ray\\ 
\hline
&&&&\\
\Mrk&0&\IRAS$^{*}$&Gonz\'alez-Delgado \& P\'erez 1996&this work\\
IRAS 13349+2438&+2&Beichman \etal 1986&Wills \etal 1992&Brandt \etal
1996\\
MCG-6-30-15&-2&\IRAS$^{*}$&Morris \& Ward 1988&Walter \& Fink 1993\\
Akn 564&-7&-&Cruz-Gonz\'alez \etal 1994&Brandt
\etal 1994\\
NGC4051&-5&\IRAS$^{*}$&Ho, Filippenko \& Sargent 1995&McHardy
\etal 1995\\
\hline
\multicolumn{5}{l}{$^{*}$
\IRAS fluxes were obtained via
the
NASA extragalactic database (NED).}\\
\multicolumn{5}{l}{All ultraviolet data obtained from the Rutherford
Appleton IUE archive.}
\end{tabular}
\end{table*}

The outlying objects in the Walter \& Fink sample are also very similar 
to \Mrk\  in their X--ray spectral and 
variability properties.
NGC 4051 has a highly variable power law component which softens as it
increases in flux (Papadakis \& Lawrence 1995) 
and a soft excess blackbody component with kT \(\sim 100\)
eV (Mihara \etal 1994, Pounds \etal 1994). Papadakis \& Lawrence's
cross correlation between the
soft (0.1 - 2 keV) and hard (2 - 8 keV) flux is 
highly asymmetric such that the soft flux tends to
lead the hard flux, and the power spectrum of the soft flux is 
steeper than that of the hard flux (\ie the soft flux has more change over  
longer time scales) which would be consistent with a slowly varying soft
excess and a rapidly varying power law component.
Akn 564 shows variability by around 20\% over 1500s during the \ros\
observations of Brandt \etal (1994), 
and when the soft
excess is modeled as a blackbody  the power law component has a similar
slope (\(\Gamma \sim 2.1\)) and the blackbody a similar temperature (kT
\(\sim 130\) eV) to \Mrk.
IRAS 13349+2438 is also variable, although as yet no rapid variability has
been seen,
and when the \ros\ spectrum is 
modeled as a power law and blackbody soft excess has a
similar power law index (\(\Gamma \sim 1.8\)) and soft excess temperature
(kT \(\sim\) 90 eV) to \Mrk, (Brandt \etal
1995).
MCG-6-30-15 also varies rapidly (Reynolds \etal 1995), and has a hard
power law component which softens as it increases in flux, although no soft
excess has as yet been detected.

Brandt \etal (1994 and 1995) found no evidence for any intrinsic neutral
column in \ros\ spectra of Akn 564 or IRAS 13349+2438, and preferred spectral
models without a soft excess component because the presence of a soft excess
would require a neutral
column which is smaller than the measured Galactic column; their
spectral fitting was based on \ros\ observations performed in December 1992
(IRAS 13349+2438) and November 1993 (Akn 564), with both targets at the
centre of the PSPC. As discussed in Section  \ref{sec:comparing} and
Appendix A, 
\ros\ observations made around these times with the target at
the centre of the PSPC were particularly affected by the PSPC temporal/spatial
gain variation. Without the recalibration software now available, the most
marked effect is that the  
fitted absorbing column is underestimated.
Even with an edge to represent absorption from O VII and O VIII,
the underlying spectra in these two objects are unusually steep
(\(\Gamma>2.5\)), suggesting
that they {\it do} have soft excesses. 

The very similar multiwavelength spectra and X--ray properties in these
objects probably mean that they lack  ultraviolet flux
for a common reason. 

Attributing the lack of ultraviolet flux to reddening by dust in
the line of sight is an attractive
solution. Optical polarization of \Mrk, of \(\sim 2\%\) increasing
to the blue, indicates scattering from dust grains. These dust grains are
probably located within the narrow line region, because the broad lines show
more 
polarization than the narrow lines (Goodrich 1989). 
For \Mrk, the required amount of reddening to make the ultraviolet
flux match the Walter \& Fink (1993) ultraviolet -- soft X--ray relation 
is consistent with  its
high Balmer decrement (\(H_{\alpha}/H_{\beta}=5.1\)) seen in the optical
spectrum. 
This amount of reddening is about twenty times that expected for a Galactic
gas to dust ratio and the measured cold absorbing column of \(\sim 
3 \times 10^{20}{\rm cm^{-2}}\) (Walter \& Fink 1993). This is reasonable if
the dust is located in the narrow line region where hydrogen and helium will
be predominantly ionized. The cross section for photoelectric absorption by
metals in dust grains can be much smaller than for metals in the gas phase
because of self blanketing
(\eg Fireman 1974 states that the 0.3 keV photoelectric cross section for 
0.6 \(\mu\) radius dust grains is  only \(\sim 20\%\) its value
for gaseous metals.)
Similar conclusions can be drawn from 
polarization properties and Balmer decrements 
of IRAS 13349+2438 and MCG-6-30-15 (Wills \etal 1992,
Thomson \& Martin 1988). 

Thermal emission from 
dust is also probably the best explanation of the
large \IRAS bumps in these objects. 
A large excess of 10 - 100 micron emission appears to be a common property
in many NLS1 galaxies, including those which are ultrasoft
such as RE J1034+396 (Puchnarewicz \etal 1995). 
Detailed sub-millimetre observations of
the NLS1 1Zw1 indicate that the IRAS bump is well fitted by
thermal emission from dust, but is poorly fitted by synchrotron models
(Hughes, \etal 1993).
The evidence for dust in these objects is extremely strong, {\it even if}
the deficit of
ultraviolet flux is related to unusually high big blue bump temperatures.

It is therefore likely that dusty Seyfert 1 galaxies are preferentially 
those which have
narrow lines. This is supported by the fact that the 
only five Seyfert 1s of the Walter \&
Fink (1993) sample to show evidence for dust reddening have narrow
lines. It is further supported by the large ratio of narrow to
broad line Seyfert 1s found in \IRAS surveys (\(>\) 20\% 
in the \IRAS samples of
Spinoglio \& Malkan 1989 and Osterbrock \& DeRobertis 1985) and small ratio
(\(\sim 10\%\)) of narrow to broad line Seyfert 1s found in optical surveys
(Stephens 1989). 

Finally, in the unified model for AGN, Seyfert 2
galaxies are Seyfert 1 galaxies viewed side-on, such that the broad line
regions are obscured from us.
The hidden broad line regions (and Seyfert 1 type optical continua) 
of Seyfert 2 galaxies can be seen in polarized
light. This is scattered into our line of sight by free electrons,
which scatter optical light more or less independently of wavelength, or
dust, which scatters blue light much more than red light (Antonucci 1993).
If more dust is found beyond the broad line region in NLS1s 
than in broad line Seyfert 1s, we would expect those Seyfert 2s which have
the largest red-blue polarization gradient to have hidden broad line
profiles like those of NLS1s. The numbers and qualities of
spectropolarimetric observations of Seyfert 2 galaxies are not sufficient to
address this issue quantatively, but this hypothesis is qualitatively 
supported: 
of the Miller \&
Goodrich (1990) sample of eight highly polarized Seyfert 2 galaxies, one
(NGC7674) has
NLS1 width polarized broad \half and \hbeta components, and this object
also has the highest red-blue polarization gradient.

\section{Conclusions}
\label{sec:conclusions}

We have examined the variable 
soft X--ray emission of \Mrk\  using hardness ratios and spectral analysis of 
9 \ros\ datasets.
The spectrum is well described by a power law and a blackbody 
(kT \(\sim 70\) eV)
soft excess. Modeling of 31 spectra at different count rates and from different
observations is consistent with a variable power law component, which is softer
when it is brighter, and an
unchanging black body soft excess.
Hardness ratios and variability amplitudes in three energy bands are in
agreement with this picture, showing
that the power law component varies continuously on a time scale of
\(\sim 5000\)s but the soft excess component probably does not vary
significantly within the observations. 
This
variability can be explained if the power law is produced by thermal or
non-thermal Comptonization of soft photons. While pair reprocessing could 
produce the correlation between power law spectral index and flux, 
delays expected between changes in the hard and
soft flux are not seen.

The behaviour of \Mrk\  is highly analogous to that of Galactic black hole
candidates in the low state, which have rapidly variable power law components 
and less variable (extended) soft excesses.

The time scale
for variation in the \ros\ band probably provides an upper limit for the size
of the region in which the power law (but not necessarily the soft excess) 
is produced. This means that the physical size of the soft excess region may
be much larger
than the X--ray power law producing region, and hence is consistent with the
soft excess having similar variability time scales to the ultraviolet part
of the big blue bump.

\Mrk\  is known to have a deficit of ultraviolet flux relative to its X--ray
emission. We show that the four other objects
in the Walter \& Fink (1993) sample of \ros\ all sky survey selected Seyfert
1 galaxies, which show a lack of ultraviolet emission (excluding IC4329a
which is probably strongly absorbed by its edge-on host galaxy) 
have very similar infrared to X--ray flux distributions to \Mrk. These
galaxies could all be classed as NLS1s, and are seen to have very similar
X--ray spectral and variability properties to \Mrk\  (calibration
uncertainties in the \ros\ PSPC mean that two of these objects, IRAS
13349+2438 and Ark 564, may be more like \Mrk\  than was previously thought). 
We argue that the multiwavelength flux distributions of these objects support
the Walter \& Fink suggestion that the ultraviolet flux deficit is related
to dust extinction, and show that other independent data 
are consistent with a picture
in which NLS1s have more dust in their narrow line regions 
than broad line Seyfert 1s.

\section{Acknowledgments}

MJP acknowledges the support of a PPARC studentship throughout much of
this work. Partial financial support for FJC was provided by the DGES under
project PB95-0122. This research has made use of data obtained from the
Leicester Database and Archive Service at the Department of
Physics and Astronomy, Leicester University, UK. This research has made use
of the NASA/IPAC Extragalactic Database (NED)   
which is operated by the Jet Propulsion Laboratory, California Institute   
of Technology, under contract with the National Aeronautics and Space  
Administration.

\section{References}

\refer Antonucci R., 1993, Annu. Rev. Astron. Astrophys., 31, 473

\refer Beichman C.A., Soifer B.T., Helou G., Chester T.J., Neugebauer G.,
Gillett F.C., Low F.J., 1986, ApJ, 308, L1

\refer Boller Th., Brandt W.N. and Fink H.H., 1996 A\&A, 305, 53

\refer Boller Th., Tr\"umper J., Molendi S., Fink H., Schaeidt S., Caulet
A., Dennefeld M., 1993, A\&A, 279, 53

\refer Brandt W.N., Fabian A.C., Nandra K., Reynolds C.S., Brinkmann W.,
1994, MNRAS, 271, 958

\refer Brandt W.N., Fabian A.C. and Pounds K.A., 1996, MNRAS, 278, 326

\refer Brandt W.N., Pounds K.A., Fink H., 1995, MNRAS, 273, L47

%\refer Breeveld A.A., Puchnarewicz E.M., 1998, MNRAS, in press

\refer Cruz-Gonzalez I., Carrasco L., Serrano A., Guichard J., 
Dultzin-Hacyan D., Bisiacchi G. F., 1994, ApJS, 94, 47

\refer Done C., Fabian A.C., 1989, MNRAS, 240, 81

\refer Done C., Fabian A.C., Ward M.J. Kunieda H., Tsuruta S., 1990,
MNRAS, 243, 713

\refer Done C., Madejski M., Mushotzky R.F., Turner T.J., Koyama K., Kunieda
H., 1992a, ApJ, 400, 138

\refer Done C., Mulchaey J.S., Mushotzky R.F., Arnaud K.A., 1992b, ApJ, 395,
275

\refer Ebisawa K., Titarchuk L., Chakrabarti S.K., 1996, PASJ, 48, 59

\refer Edelson R.A. and Krolik J.H., 1988, ApJ, 333, 646

\refer Edelson R.A., 1992, ApJ, 401, 516

\refer Fireman E.L., 1974, ApJ, 187, 57

\refer Forster K., Halpern J.P., 1996, ApJ, 468, 565

\refer Gonz\/alez Delgado R.M. \& P\/erez E., 1996, MNRAS, 278, 737

\refer Goodridge R.W., 1989, ApJ, 342, 224

\refer Grupe D., Beuerman K., Mannheim K., Thomas H.-C., Fink H.H.,
de Martino D., 1995, A\&A, 300, L21

\refer Guilbert P.W. and Rees M.J., 1988, MNRAS, 233, 475

\refer Hasinger G., Boese G., Predehl P., Turner T.J., Yusaf R.,
George I.M., Rohrbach G., 1994a, MPE/OGIP Calibration Memo CAL/ROS/93-015 

\refer Hasinger G.H., Prieto A., Snowden S., 1994b, MPE \ros\ Newsletter issue 10,

\refer Ho L.C., Filippenko A.V. \& Sargent W.L., 1995, ApJS, 98, 477

\refer Hughes D.H., Robson E.I., Dunlop J.S., Gear W.K., 1993, MNRAS, 263, 607

\refer Kunieda H., Hayakawa S., Tawara Y., Koyama K., Tsusaka Y., Leighly
K., 1992, ApJ, 384, 482

\refer Lawson A.J., Turner M.J.L., Williams O.R., Stewart G.C.,
Saxton R.D., 1992, MNRAS, 259, 743

\refer Leighly K.M., Mushotzky R.F., Yaqoob T., Kunieda H., Edelson R., 
1996, ApJ, 469, 147

\refer Matsuoka M., Piro M., Yamauchi M., Murakami T., 1990, ApJ, 361, 440

\refer McHardy I.M., Green A.R., Done C., Puchnarewicz E.M., Mason K.O.,
Branduardi-Raymont G., Jones M.H., 1995, MNRAS, 273, 549

\refer Miller J.S. \& Goodrich R., 1990, ApJ, 355, 456

\refer Mihara T., Matsuoka M., Mushotzky R.F., Kunieda H., Otani C.,
Miyamoto S., Yamauchi M., 1994, PASJ, 46, 137

\refer Miyamoto S. \& Kitamoto S., 1989, Nature, 342, 773

\refer Molendi S., Maccacaro T. and Schaeidt S., 1993, A\&A, 271, 18

\refer Molendi S. and Maccacaro T., 1994, A\&A, 291, 420

\refer Morris S.L. \& Ward M.J., 1988, MNRAS, 230, 639

\refer Nandra K., George I.M., Mushotzky R.F., Turner T.J., Yaqoob T., 1997,
ApJ, 477, 602

\refer Netzer H., Turner T.J. and George I.M., 1994, ApJ, 435, 106

\refer Osterbrock D.E. \& DeRobertis M.M., 1985, PASP, 97, 902

\refer Otani C., Tsuneo K., Kayoko M., 1996, MPE Report 263,
`Rontgenstrahlung from the Universe', Zimmermann U., Trumper J.E., Yorke H.
(eds)

\refer Papadakis I.E., \& Lawrence A., 1995, MNRAS, 272, 161

\refer Pounds K.A., Done C., Osborne J.P., 1995, MNRAS, 277, 5P

\refer Pounds K.A., Nandra K., Fink H.H. and Makino F., 1994, MNRAS, 267, 193

\refer Pounds K.A., Nandra K., Stewart G.C., George I.M., Fabian A.C.,
1990, Nature, 344, 132

\refer Pounds K.A., Turner T.J., Warwick R.S., 1986, MNRAS, 221, 7P

\refer Prieto M.A., Hasinger G., Snowden S., 1994, MPE Calibration Memo
TN-ROS-MPE-ZA00/032

\refer Puchnarewicz E.M., Mason K.O., C\/ordova F.A., Kartje J.,
Branduardi-Raymont G., Mittaz J.P.D., Murdin P.G., Allington-Smith J., 1992,
MNRAS, 256, 589

\refer Puchnarewicz E.M., Mason K.O., Siemiginowska A., Pounds K.A., 1995,
MNRAS, 276, 1281

\refer Reynolds C.S., Fabian A.C., Nandra K., Inoue H., Kunieda H., Iwasawa
K., 1995, MNRAS, 277, 901

\refer Ross R.R., Fabian A.C. and Mineshige S., 1992, MNRAS, 258, 189

\refer Ross R.R. and Fabian A.C., 1993, MNRAS, 261, 74

\refer Shastri P., Wilkes B.J., Elvis M., McDowell J., 1993, ApJ,
410, 29

\refer Snowden S.L., McCammon D., Burrows D.N. and Mendenhall J.A., 1994, ApJ,
424, 714

\refer Snowden S.L., Turner T.J., George I.M., Yusaf R., 
Predehl P., Prieto A., 1995, OGIP Calibration Memo CAL/ROS/95-003

\refer Spinoglio L. \& Malkan M.A., 1989, ApJ, 342, 83

\refer Stark A.A., Gammie C.F., Wilson R.W., Bally J., Linke R., Heiles C.,
Hurwitz M., 1992, ApJS, 79, 77

\refer Stephens A.J., 1989, 97, 10

\refer Svensson R., 1987, MNRAS, 227, 403

%\refer Shakura N., Sunyaev R., 1973, A\&A, 24, 337

\refer Tanaka Y., Nandra K., Fabian A.C., Inoue H., Otani C., Dotani
T., Hayashida K., Iwasawa K., Kii T., Kunieda H.,
Makino F., Matsuoka M., 1995, Nature, 375, 659 

\refer Thompson I.B., Martin P.G., 1988, ApJ, 330, 121

\refer Torricelli-Ciamponi G. \& Courvoisier T.J.-L., 1995, A\&A, 296, 651

\refer Turner T.J., 1993, OGIP Calibration Memo CAL/ROS/93-007

\refer Turner T.J., Pounds K.A., 1989, MNRAS, 240, 833

\refer Walter R., \& Courvoisier T.J.-L., 1992, A\&A, 266, 65

\refer Walter R. \& Fink H.H., 1993, A\&A, 274, 105

\refer Wills B.J., Wills D., Evans N.J., Natta A., Thompson K.L., Breger M.,
Sitko M.L., 1992, ApJ, 400, 96

\refer Zdiarzki A.A., Ghisellini G., George I.M., Svensson R., Fabian A.C.,
Done C., 1990, ApJ, 363, L1

\vspace{1cm}
\begin{boldmath}
\appendix{\bf \noindent
 A\ \ \  The $ROSAT$ PSPC gain drift and the \Mrk\ data}
\end{boldmath}
\vspace{3mm}

In Section \ref{sec:comparing} we stated that observations P4, P8 and P9 are
probably most affected by the PSPC spatial/temporal gain drift. 
It {\it is} 
known that there has been a significant change in response 
{\it on-axis} between the beginning and end of the PSPC lifetime (Snowden
\etal 1995). The latest information (M.J. Freyberg \etal
1996, {\it http://wave.xray.mpe.mpg.de/rosat\- 
/calibration/pspc/PSPC\_cal\_status.html}) 
is that large residuals are found in bright,
on-axis sources observed later than mid 1992. From this one would expect
observations P4, P8 and P9 to be affected by the gain drift; it is unclear
whether observation P2 should be affected as well.
Software to correct for this gain variation is now available in the form of the
PCPICOR task in FTOOLS and PROCESS/CT in EXSAS, but because the investigation
reported here was begun before this software was generally available, it was
initially performed without this correction. The difference between the
corrected and uncorrected spectra is considerable.

The most noticeable difference concerns the soft end of the 
P4, P8 and P9 spectra: before correction with PCPICOR 
they all had smaller fitted absorption and 
higher black body
temperatures, (\eg for comparison with model B in Table \ref{tab:fitting}, P4,
P8 and P9 had best fit columns of 1.7, 1.8 and 2.0 \(\times 10^{20} {\rm
cm^{-2}}\) and best fit black body temperatures of 99, 87 and 105 eV
respectively).
As an illustration of this difference we show the pre and post PCPICOR
residuals of a power law fit assuming Galactic \(\nH\) to the P9 data in Fig. 
A1
and Fig. A2. Note how different the pre-correction residuals are to those for
the P2 observation shown earlier (Fig. \ref{fig:tmprati}).
The offaxis observation P5 was affected in the opposite sense by PCPICOR, in
that the fitted column decreased after correction.

However, despite the improvement that PCPICOR makes, the three on-axis
observations {\it still} have lower fitted columns and higher fitted
temperatures than the most of the other observations, suggesting they may still
have calibration differences which are significant in data of this signal to
noise. The scatter of fitted
temperatures and columns in Table \ref{tab:fitting} is of the same order as the
differences made by PCPICOR; we therefore consider that the different
fitted temperatures and columns for the different observations could be due to
remaining PSPC calibration uncertainties.

\begin{figure}
\begin{center}
\leavevmode
\psfig{file=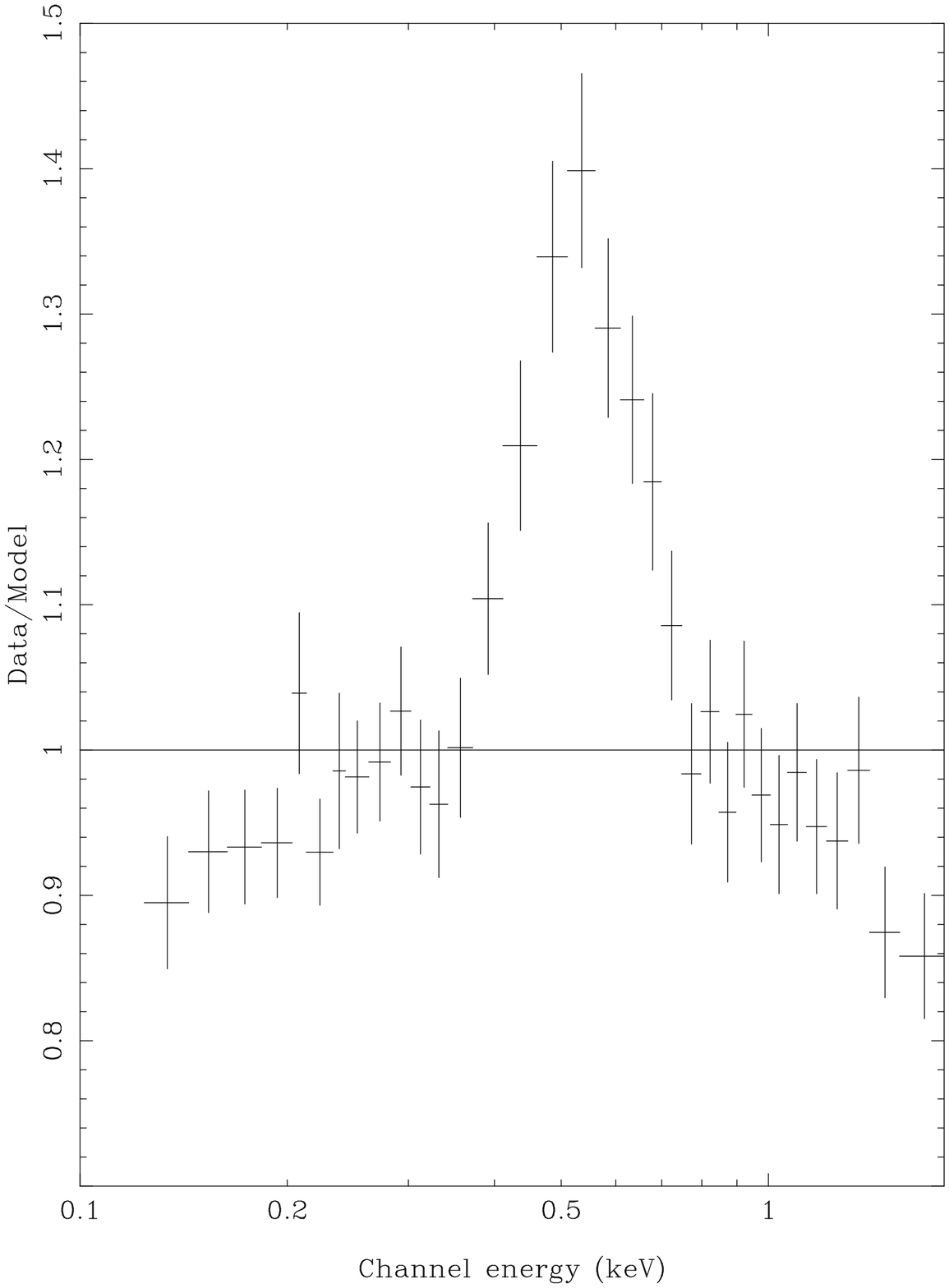,width=85mm,height=110mm}
{\bf Figure A1} Ratio of a power law model (best fit $\Gamma=2.2$)
with fixed Galactic $\nH$,
to the P9 spectrum {\it before} correction using PCPICOR. 
Note that the residuals are found at higher energies
than for the P2 spectrum shown in Fig. \ref{fig:tmprati}.
\label{fig:old91arati}
\end{center}
\end{figure}

\begin{figure}
\begin{center}
\leavevmode
\psfig{file=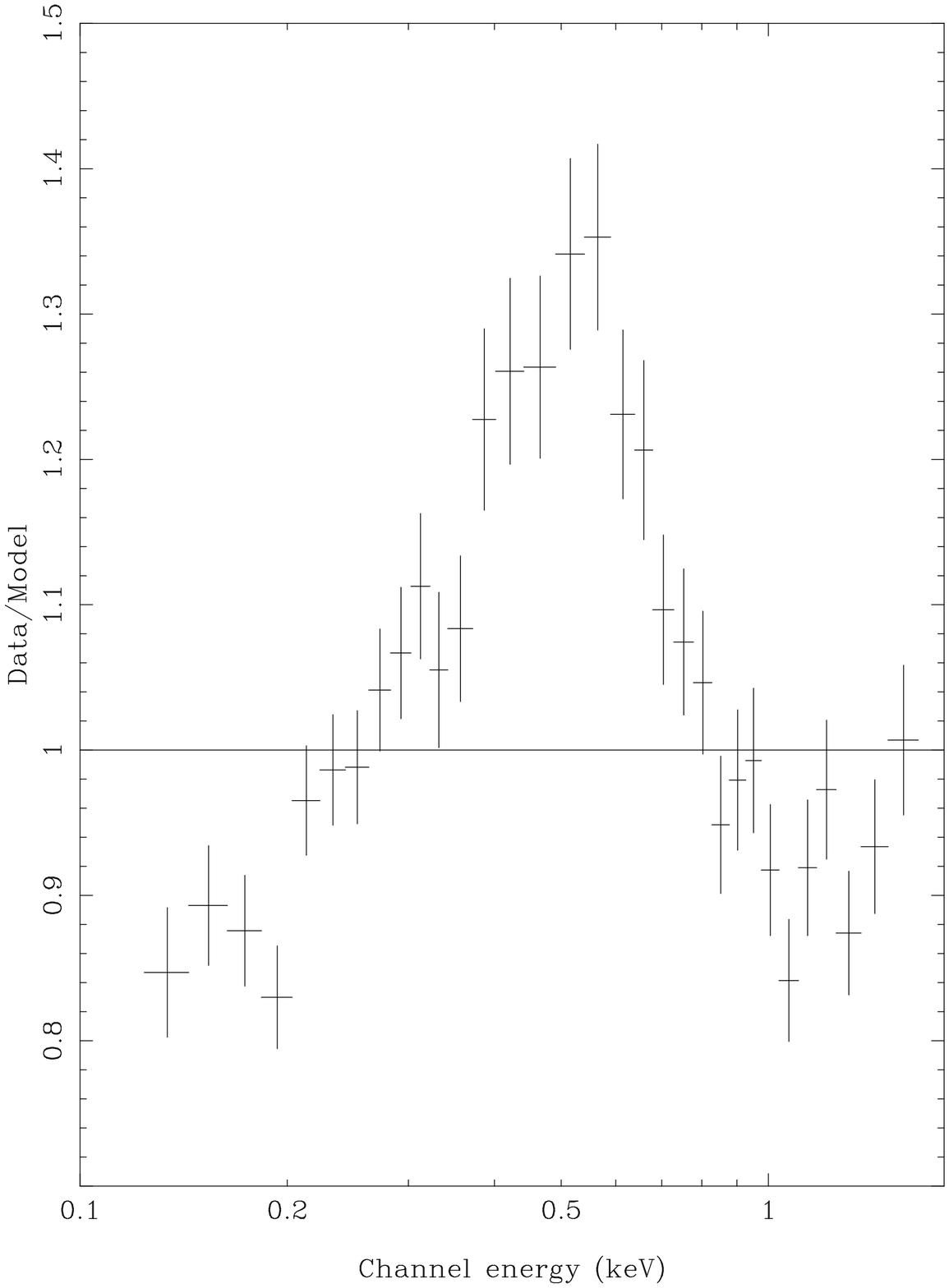,width=85mm,height=110mm}
{\bf Figure A2} Ratio of a power law model (best fit $\Gamma=2.2$)
with fixed Galactic $\nH$,
to the P9 spectrum {\it after} correction with PCPICOR. 
Note that the residuals now have a similar shape to those in Fig. 
\ref{fig:tmprati}
\label{fig:new91arati}
\end{center}
\end{figure}

\end{document}